%% file: paper.tex
\documentclass[twocolumn]{aastex631}

\usepackage[super]{nth}
\usepackage{savesym}
\savesymbol{tablenum}
\usepackage[T1]{fontenc}
\usepackage{inputenc}
\usepackage[range-units = brackets,tophrase={-},seperr,repeatunits=false]{siunitx}
\usepackage{multirow}
\restoresymbol{SIX}{tablenum}

\newcommand{\kms}{\si{km\,s^{-1}}}

\graphicspath{{./}{figures/}}

\accepted{\today}

\submitjournal{ApJ}

\shorttitle{Feedback in the merging galaxy group NGC~6338}
\shortauthors{Schellenberger et al.}

\begin{document}

\title{Feedback in the extremely violent group merger NGC~6338}
\correspondingauthor{Gerrit Schellenberger}
\email{gerrit.schellenberger@cfa.harvard.edu}

\author[0000-0002-4962-0740]{Gerrit Schellenberger}
\affiliation{Center for Astrophysics $|$ Harvard \& Smithsonian, 60 Garden St., Cambridge, MA 02138, USA}

\author[0000-0002-5671-6900]{Ewan O'Sullivan}
\affiliation{Center for Astrophysics $|$ Harvard \& Smithsonian, 60 Garden St., Cambridge, MA 02138, USA}

\author[0000-0002-1634-9886]{Simona Giacintucci}
\affiliation{Naval Research Laboratory, 4555 Overlook Avenue SW, Code 7213, Washington, DC 20375, USA}

\author{Jan Vrtilek}
\affiliation{Center for Astrophysics $|$ Harvard \& Smithsonian, 60 Garden St., Cambridge, MA 02138, USA}

\author{Laurence P. David}
\affiliation{Center for Astrophysics $|$ Harvard \& Smithsonian, 60 Garden St., Cambridge, MA 02138, USA}

\author[0000-0003-2658-7893]{Francoise Combes}
\affiliation{Observatoire de Paris, LERMA, Coll\`ege de France, PSL Univ., CNRS, Sorbonne Univ.,  Paris, France}

\author{Laura B{\^i}rzan}
\affiliation{Hamburger Sternwarte, Universit{\"a}t Hamburg, Gojenbergsweg 112, 21029 Hamburg, Germany}

\author[0000-0002-1370-6964]{Hsi-An Pan}
\affiliation{Department of Physics, Tamkang University, No.151, Yingzhuan Road, Tamsui District, New Taipei City 251301, Taiwan}

\author[0000-0001-7218-7407]{Lihwai Lin}
\affiliation{Institute of Astronomy and Astrophysics, Academia Sinica, No.  1, Section 4, Roosevelt Road, Taipei 10617, Taiwan}

\begin{abstract}
The galaxy group NGC~6338 is one of the most violent group-group mergers known to date. 
While the central dominant galaxies rush at each other at $1400\,\kms$ along the line of sight, with dramatic gas heating and shock fronts detected, the central gas in the BCGs remains cool. There are also indications of feedback from active galactic nuclei (AGNs), and neither subcluster core has been disrupted.  

With our deep radio uGMRT data at 383\,MHz and 650\,MHz we clearly detect a set of large, old lobes in the southern BCG coinciding with the X-ray cavities, while the northern, and smaller BCG appears slightly extended in the radio.
The southern BCG also hosts a smaller younger set of lobes, perpendicular to the larger lobes, but also coinciding with the inner X-ray cavities, and matching the jet direction in the parsec-resolution VLBA image. Our spectral analysis confirms the history of two feedback cycles. The high radio frequency analysis classifies the compact source in the southern BCG with a powerlaw, while ruling out a significant contribution from accretion. The radio lightcurve over 3 decades shows a change about 10 years ago, which might be related to ongoing feedback in the core. 

The southern BCG in the NGC~6338 merger remains another prominent case where the direction of jet-mode feedback between two cycles changed dramatically.

\end{abstract}


\section{Introduction} \label{sec:intro}
The standard $\Lambda$CDM cosmological model predicts collapsed structures growing from small density fluctuations through gravity. These collapsed overdensities can become the largest objects in the Universe through the infall and merger of surrounding matter. Consequently, galaxy clusters, the largest gravitationally bound structures, are undergoing mergers at multiple stages during their evolution. Galaxy cluster mergers are the most energetic events in the Universe ($\SI{e63}{erg\,s^{-1}}$), with expected merger velocities around 500 to $\SI{700}{\kms}$ based on cosmological simulations (\citealp{Thompson2012-fa}). Observations of mergers with much higher velocities (\citealp{Markevitch2002-fl}) require a tail to this velocity distribution. 

Galaxy clusters are filled with a hot, X-ray bright plasma (intra cluster medium, ICM), with central cooling timescales often less than $\SI{1}{Gyr}$, predicting large amounts of cold molecular gas in the centers of clusters that can fuel star formation (\citealp{Fabian1984-kj,Mark_Voit2011-lx,Stern2019-ph}).
However, the observed amounts of cold gas and star formation are largely far below the high predictions. 
A solution began to appear after the discovery of powerful relativistic jets launched by the supermassive black holes (SMBHs) in the cluster-center galaxies heating the ICM (\citealp{Churazov2001-az}). It is now established that active galactic nuclei (AGNs, which harbor the SMBH) are part of the feedback mechanism that regulates the balance between infalling cold gas that has condensed from the ICM, and jets mechanically heating the environment (e.g., \citealp{McNamara2000-pl,Forman2007-ew}). 
The energy available from the jets can be estimated from X-ray measurements of the volume and pressure of the displaced ICM. \cite{Birzan2004-pw,OSullivan2011-xb,Panagoulia2014-tg} and others have shown that the energy output is sufficient to prevent most of the ICM from cooling. 

The released energy of a major merger can significantly disturb the feedback cycle that is established within a galaxy cluster or group, and even disrupt cool cores. 
Observationally, the correlation between cool core / non-cool core and merging state has been demonstrated (e.g., \citealp{Chon2012-oq,Lovisari2017-im}). 
On the simulation side, several studies confirm the observed trend  (e.g., \citealp{ZuHone2011-lw,Valdarnini2021-tv}) and point out that cool cores do not survive a major merger due to the high entropy gas brought into the core. Only off-axis mergers with smaller mass ratios tend to retain the cool core properties. However, some other simulation studies do not find a clear trend of cool core clusters being more relaxed (e.g., \citealp{Barnes2018-wp}), and indicate that events other than mergers can disrupt cool cores. More recently, \cite{Sharma2021-fx} found that galaxy simulations of gas rich major mergers do not enhance the AGN activity, as one might expect due to the new supply of cold gas. 
Therefore, to understand the AGN feedback process fully, and how it might be triggered or interrupted through mergers, observations of merging systems with an ongoing feedback cycle are particularly interesting.

Despite most studies focusing on galaxy clusters, galaxy groups turn out to be at least as important: They host most of the galaxies (\citealp{Eke2006-ld}) and matter in the universe (\citealp{Fukugita1998-bi}). 
Their hot intra-group medium (IGM) cools even faster than the ICM through X-ray line emission, while the gravitational potential is shallower, which requires strong and fine-tuned AGN feedback as a balance. 
Yet their dominant galaxies show little star formation (SF) and only limited cold gas content (\citealp{OSullivan2018-yk,McDonald2018-xg,Schellenberger2020-vl,Kolokythas2022-te}). 
The possibility has been proposed that gas has been driven out to large radii by the AGN, diminishing the efficiency of AGN re-heating (e.g., \citealp{Eckert2021-mm}). At the same time, star formation is also highly suppressed in the central dominant galaxies (CDGs) of groups with respect to CDGs of clusters. 

The nearby galaxy group NGC741 shows many of these interesting features: \cite{Schellenberger2017-zl} describe the narrow X-ray filaments around the old radio galaxy in the center of the group (also called NGC741), and the vicinity of the infalling head-tail radio galaxy, NGC742, which has strongly bent jets and a radio tail extending more than 100\,kpc in projection. The ongoing merger shows signs of disturbance in the group (sloshing, rings in the optical image of the infalling galaxy), but the cool core and the structure of the BCG seem to be largely unaffected. 

A far more massive merger has been observed in the galaxy group NGC~6338 (\citealp{Dupke2013-lm,Wang2019-xz,OSullivan2019-le}). It appears to be a unique case of a highly violent group-group merger with velocities up to $1800\,\kms$, where dominant galaxies show signs of feedback. 
The balancing of the feedback cycle during an ongoing major merger is largely unknown and can be studied in detail in NGC~6338. 
The hydrostatic mass from the X-rays is close to $\SI{e14}{M_\odot}$, and past studies have found that the merger interaction heats the gas up to 5\,keV and create shocks in the ICM. The merger is mostly along the line of sight, with the larger dominant galaxy located to the south. The two dominant galaxies are separated in projection by about 1\,arcmin (32\,kpc). 
\cite{Pan2020-ei} report H$\alpha$ and CO detections in the northern core, which both appear slightly offset from the center of the galaxy. The cold and warm gas masses are consistent with cooling predictions, implying an offset cooling scenario in the northern BCG.  

The southern BCG contains bright, colder X-ray gas with short cooling times, H$\alpha$ filaments that extend 17\,arcsec (8\,kpc in projection) and are correlated with the X-ray structures, and hosts an AGN that is visible in the radio band, which indicates ongoing feedback. 
Its stellar mass, $\log M_{\star, {\rm BCG}}=11.47$ (\citealp{Marino2016-im}), is relatively high for a group merger constituent, with a rotation around the major photometric axis (\citealp{Tsatsi2017-qt}). 
Star formation estimates from K-band measurements show a large value around $\SI{1}{M_\odot\,yr^{-1}}$, but the presence of the AGN might lead to an overestimate (\citealp{ODea2008-jy}).
More detailed IR photometry (\citealp{Crawford1999-cw,Quillen2008-ap}) classifies the BCG as quiescent and indicates the presence of lines such as OIII. 

We recently obtained deep uGMRT observations in two bands to answer the open question: Are there radio jets or radio lobes in one or both of the cores. These observations will help to understand the feedback history, and compare it to relaxed galaxy groups with strong feedback, such as NGC\,5044 , which is the X-ray brightest group in the sky. NGC\,5044 shows strong signs of gas cooling throughout the waveband (cold X-ray and H$\alpha$ filaments described in \citealp{David2017-ig}, [CII] and [NII] line emission and CO emission; \citealp{David2014-jn,Werner2014-vw}, old radio plasma; \citealp{OSullivan2013-ef}). Three distinct AGN cycles are imprinted in the hot X-ray bright gas, and a new feedback cycle of the AGN is just starting (\citealp{Schellenberger2020-ji}).

This paper is structured as follows. In Section \ref{ch:observations}, we describe the  
setup and data reduction of the recent uGMRT observations. Several other datasets have been included in this study, such as archival VLA and VLBA observations, and our recent Submillimeter Array (SMA\footnote{The Submillimeter Array is a joint project between the Smithsonian Astrophysical Observatory and the Academia Sinica Institute of Astronomy and Astrophysics and is funded by the Smithsonian Institution and the Academia Sinica.}) and IRAM observation at high frequencies, which are also described in section \ref{ch:observations}. Our results, such as the description of past feedback cycles in the southern BCG, and the current state of the AGN and connected jets, are presented in section \ref{ch:results}. Section \ref{ch:discussion} provides a discussion of the radio and sub-mm results, including the age of past feedback cycles, and opportunities for future X-ray missions. We summarize our findings in section \ref{ch:summary}.

Throughout this paper we assume a flat $\Lambda$CDM cosmology with the following parameters: $\Omega_{\rm m} = \num{0.3}$, $\Omega_\Lambda = \num{0.7}$, $H_0 =  h \cdot \SI{100}{km~s^{-1}~Mpc^{-1}}$ with $h = \num{0.7}$, which gives an angular scale of $\SI{0.528}{kpc}$ per arcsec at redshift 0.027 ($D_A=\SI{109}{Mpc}$) of NGC~6338. 
Uncertainties are stated at the 68\% confidence level, and cluster masses refer to a radius within which the mean density of the cluster is 500 times the critical density of the Universe at the cluster redshift.

\section{Observations}
\label{ch:observations}
The main goal of the dedicated uGMRT observations was to find signatures of feedback, which is ideal for this instrument with its decent spatial resolution at low frequencies (below $\sim \SI{1}{GHz}$). We decided to expand this by including higher frequency data from the VLA and several other published values to construct and fit an SED over 3 orders of magnitude in frequency, which reveals the emission mechanisms. At the high frequency end we included our recent SMA continuum observation at  $\SI{230}{GHz}$ to complete the SED. Repeated VLA observations indicate the radio variability of the central radio source in the southern BCG, and the IRAM 30m CO measurements can quantify the reservoir of cold molecular gas. 

\subsection{uGMRT data reduction}
\input{table_obs_gmrt}
The upgraded Giant Metrewave Radio telescope (uGMRT) is equipped with four wideband receivers. 
We observed NGC~6338 in bands 3 ($300-500\si{MHz}$) and 4 ($550-850\si{MHz}$)  (see Tab. \ref{tab:gmrt}). 
Both observations were recorded with 5.37s integration time, 4096 channels, and two stokes parameters. The hardware-based removal of radio frequency interference (RFI)\footnote{\url{http://www.ncra.tifr.res.in/ncra/gmrt/gmrt-users/online-rfi-filtering}}  was not enabled. 
For band 3, 27 of 30 antennas were available, while for the band 4 observation only 26 antennas were operational. Each observation ends with a 10min scan of 3C48, which was used as a flux calibrator for both datasets  (more details on the observations are given in Tab. \ref{tab:gmrt}). 
The reduction of uGMRT data utilized the SPAM pipeline (\citealp{Intema2009-cs}), and the processing is analogous to the description in \cite{Schellenberger2022-sk}. 
We note that the wideband observations were split into 6 and 5 subbands for bands 3 and 4, respectively, before the pipeline derived and applied the calibration, several self-calibration cycles, and the direction-dependent calibration to correct ionospheric disturbances. 
The quoted noise levels in Tab. \ref{tab:gmrt} are achieved after imaging the merged, calibrated data products from SPAM, using wsclean (\citealp{Offringa2014-bw}, version 2.10) with the widefield w-algorithm, a Briggs robust weighting scheme with $r=-0.25$, multiscale deconvolution, and a third order polynomial taking into account the wideband spectral behavior. 
The sources to be cleaned are detected through the pyBDSF (\citealp{Mohan2015-hz}) algorithm. We select all pyBDSF detections with at least $6\sigma$ significance. These sources are cleaned in wsclean to the $0.3\sigma$ threshold, and we visually inspected the residuals to contain only noise.  
We note that part 5 of the Band 3 observation from $\SIrange{433}{467}{MHz}$ was misaligned with respect to the other parts by 2\,arcsec after the SPAM processing. We fixed this using the CASA (\citealp{McMullin2007-ed}) \textit{fixvis} task. Other subbands did not show astrometry offsets. 

The final images have been corrected for the primary beam following the GMRT guidelines\footnote{\url{http://www.ncra.tifr.res.in/ncra/gmrt/gmrt-users/observing-help/ugmrt-primary-beam-shape}}. 
For a spectral index map between band 3 and 4 the astrometry has to be precise. We noticed a small ($<\SI{1}{arcsec}$) offset between those two bands and shifted one of the images before creating a spectral index map. We have verified the relative and absolute flux calibration of the final images in both bands by comparing the fluxes of point sources in the field. The spectral index map from two observations requires a uniform sampling of the uv-space, and identical restoring beam. Therefore, we rerun the wsclean task with some additional parameters to include only uv distances from 1k to 32k$\lambda$ (both GMRT bands sample these uv ranges well), and a circular restoring beam of $\SI{8}{\arcsec}$. This setup results in an r.m.s of $\SI{31}{\mu Jy\,bm^{-1}}$ for band 3, and $\SI{28}{\mu Jy\,bm^{-1}}$ for band 4.

\subsection{VLA data reduction}
\input{table_obs_vla}
NGC~6338 has been observed many times in the past with the VLA at higher frequencies than the GMRT (see Tab. \ref{tab:vla}). We selected observations with frequencies higher than 1\,GHz (at least L band), which can easily determine the flux of the compact radio source in the southern BCG in NGC~6338. The survey data from FIRST (\citealp{Becker1994-te}), NVSS (\citealp{Condon1998-tu}) and VLASS (\citealp{Lacy2020-kg}) have only been download from the archive -- not reprocessed, and are listed for completeness in Tab. \ref{tab:vla}. 

We reduced the archival VLA data before 2012 using the Astronomical Image Processing System (AIPS version 31DEC22, \citealp{Greisen1990-ak,Greisen2003-le}) and the \textit{vlaprocs} module. 
We load the raw VLA files into AIPS, and apply the flux scale to the corresponding flux calibrator observation with \textit{SETJY} including the options \textit{OPTYPE = CALC}. The parameter \textit{APARM(2)} is set in a way to match the time of the observation. 
We then determine the telescope-based calibration and set the flux and phase calibrator names, and determine the flux densities for the calibration sources from the primary calibrator. Finally, we apply the solutions to the target. 
Imaging of each observation has been done using \textit{tclean} in CASA with Briggs robust weighting $r=1$.

The EVLA/JVLA observations 12A182 and 15A215 have been reduced using the \textit{CASA} VLA pipeline (version 5.6.3-19), followed by imaging of the target scans with \textit{tclean} and applying several phase-only self-calibration cycles. Also these observations should only provide a flux of the radio point source. 

We note that the source flux in Tab. \ref{tab:vla} refers to the flux of a point-like source, therefore the r.m.s. value states the statistical uncertainty of the extracted fluxes. Additionally, we assume a 3\% systematic uncertainty (\citealp{Perley2017-ug}). 
Note that observation AM701 was used to characterize the high frequency radio spectrum, since all 4 bands (C, X, Ku, K) are observed simultaneously (see section \ref{ch:radio_spectrum}. We did not combine or simultaneously image VLA observations. 

\subsection{SMA data reduction}
We observed NGC~6338 at 230\,GHz with the SMA to measure the current high frequency flux (project 2021B-S068, PI Schellenberger). 
Seven of the total eight 6\,m antennas were operational during the observations performed in the compact configuration on Jan 15, 2022 and Jan 17, 2022, with total observing times of 1h 33min, and 2h 4min, respectively.  
A reliable flux calibration was ensured through a scan of Ceres during each track. The time dependence of the complex gain was calibrated through frequent observations of 1740+521. 
The total on target time sums to 139min. Data reduction was performed using the IDL package MIR\footnote{\url{https://lweb.cfa.harvard.edu/~cqi/mircook.html}} (\citealp{Gurwell2007-wz}), and the important steps include the flagging of unusable visibilities, a system temperature calibration, and a baseline-based bandpass calibration using scans of 3C345 and 3C279 with the \textit{pass\_cal} task. Using the flux calibrator observation, we constrained the gain amplitude of the phase calibrator by applying \textit{sma\_flux\_cal} and \textit{flux\_measure} for each sideband and receiver separately. For a combined imaging of both tracks in CASA  we apply the correct weights scaling factors (\textit{\mbox{MIRFITStoCASA.py}}). After two phase self-calibration steps without flagging bad solutions, we improved the noise slightly, yielding a final r.m.s. of $\SI{0.46}{mJy\,bm^{-1}}$,  and a beamsize of $6.1\times 2.7\,\si{arcsec}$. 
The measured flux at 225\,GHz of the central point source with the SMA is $\SI{3.9(5)}{mJy}$. 

\subsection{VLBA data reduction}
To better understand the current state of the southern BCG in NGC~6338 we analyzed the archival VLBA observation (Project BE063/F, PI A. Edge) in C band taken in May 2013. The scan of NGC~6338 has data taken from 9 antennas (SC missing), and a spectral setup of 8 spectral windows with each having a 32MHz bandwidth. 

We used the \textit{VLBARUN} task in \textit{AIPS} to perform the VLBA calibration, using the scan of 3C345 as fringe finder and bandpass calibrator, and J1722+5856 for phase-referencing.
The solution interval was set to 10\,min and the selected reference antenna is Pie Town (PT).  
After exporting the 19\,min calibrated target scans we applied 5 phase-only self-cal cycles, reaching an r.m.s. $\SI{74}{\mu Jy\,bm^{-1}}$ and a restoring beam of $4.0\times 1.1\si{mas}$. 
The integrated source flux yields $\SI{26}{mJy}$ at $\SI{5}{GHz}$. 
Since the observation is only a short snapshot, the uv-plane is not sufficiently filled for the longest baselines. To avoid any bias in determining the jet direction, we limit the final image baselines below $\SI{3000}{km}$, which is well sampled. The restored beam is $4.0\times 3.2\si{mas}$, and the noise level is $\SI{100}{\mu Jy\,bm^{-1}}$.

\subsection{IRAM~30m}
We observed NGC~6338 with the IRAM 30m on 2020 July 21-24 (project 066-20) using the E0 and E2 bands of the EMIR receivers and both horizontal and vertical polarizations. We observed the redshifted CO(1-0) and CO(2-1) lines simultaneously, at 112.194~GHz and 224.388~GHz respectively. Both the FTS and WILMA backends were used. Weather conditions made about half the approved time unusable, but we were able to observe the target for a total of $\sim$8 hours. The focus and pointing of the telescope were calibrated using observations of bright quasars or planets, and the data were reduced using the \textsc{gildas}\footnote{http://www.iram.fr/IRAMFR/GILDAS} \textsc{class} software \citep{Pety2005-eb,Gildas_Team2013-xb}.

\begin{figure}
    \centering
    \includegraphics[width=0.5\textwidth,trim={80px 410px 80px 200px},clip]{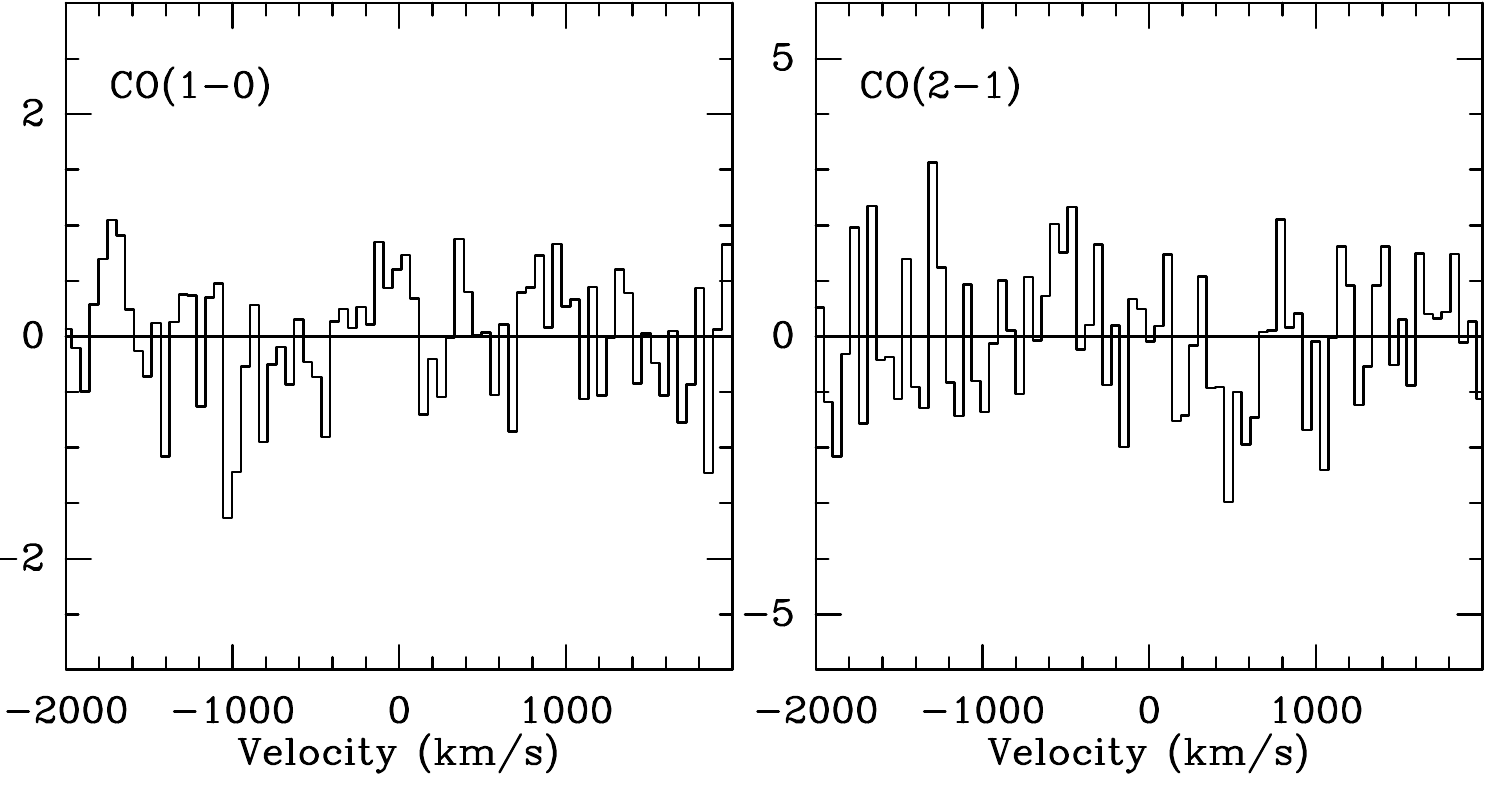}
    \caption{CO(1-0) and CO(2-1) spectra of NGC~6338 from IRAM 30m in the left and right panels respectively. The vertical axes indicate main beam temperatures (T$_{\rm mb}$) in mK, and the spectra are binned to $\SI{52}{km\,s^{-1}}$.}
    \label{fig:iram}
\end{figure}

Figure~\ref{fig:iram} shows the summed CO(1-0) and CO(2-1) spectra for NGC~6338. Although neither line was detected, we are able to place strong upper limits on the CO emission from the source. With $\SI{52}{km\,s^{-1}}$ binning, the r.m.s. antenna temperature T$_{\rm mb}$ was $\SI{0.5}{mK}$ for CO(1-0) and $\SI{1.1}{mK}$ for CO(2-1). Following the approach laid out in \citet{OSullivan2018-yk}, we adopt a nominal linewidth of $\SI{300}{km\,s^{-1}}$ as typical for a group-central galaxy, and a CO-to-H$_2$ conversion factor $\alpha_{CO}=4.6$ \citep[including helium,][]{Solomon2005-xa}. We calculate $3\sigma$ upper limits on the molecular gas mass, finding $M_{\rm mol} \leq \SI{1.4e8}{M_\odot}$ from the CO(1-0) spectrum, and 
$M_{\rm mol} \leq \SI{8e7}{M_\odot}$ from the CO(2-1) spectrum, assuming equal excitation temperature of CO(2-1) and CO(1-0), which means full local thermal equilibrium (LTE). 
However, for sub-thermal excitation with $\frac{T_{\rm CO(2-1)}}{T_{\rm CO(1-0)}} \approx 0.5$ 
we derive an upper mass threshold for CO(2-1) of $\SI{1.6e8}{M_\odot}$. 
Given the uncertainties we use $\SI{1.4e8}{M_\odot}$ as the upper limit for the molecular gas in the southern core, which is a factor 12 below the mass that would have been predicted based on the H$\alpha$ luminosity of the galaxy and the L$_{\rm H\alpha}$-$M_{\rm mol}$ relation \citep{Salome2003-ge}. 
The beam size of the IRAM~30m is $\sim \SI{22}{\arcsec}$ (HPBW) at our redshifted CO(1-0) frequency and $\sim \SI{11}{\arcsec}$(HPBW) at CO(2-1). The H$\alpha$ filaments in NGC~6338 extend to $\sim \SI{17}{\arcsec}$ radius, but the great majority of the flux is found in the central $\SI{5}{\arcsec}$. We therefore assume that all the CO flux is contained within the IRAM~30m beam.

We note that while we can only derive an upper limit for the CO (1-0) flux in the southern BCG of $\SI{0.94}{Jy\,km\,s^{-1}}$, it is comparable to the NOEMA detection of CO in the northern BCG of $\SI{1.01}{Jy\,km\,s^{-1}}$ (\citealp{Pan2020-ei}).

\section{Results}
\label{ch:results}
We describe the results obtained from the data that have been analyzed: The new uGMRT data cover low frequencies (below 1\,GHz) at  higher spatial resolution compared to existing LOFAR data ($13.7\times 8.7\,\si{arcsec}$ versus $7.9\times 5.6\,\si{arcsec}$ for the band 3 uGMRT data), allowing us to look at feedback cycles of the AGNs in the northern and southern BCG in detail. 
With the multi-frequency radio coverage from 142\,MHz to 20\,GHz the SED is parameterized by a self-consistent model, which can even be expanded to 230\,GHz. 
The archival VLBA data give a glimpse at the current state of the AGN. 

\subsection{Feedback cycles in the southern BCG}
There are strong indications that the southern BCG has established a feedback cycle that is maintained by the central AGN: \cite{OSullivan2019-le} found X-ray cavities around the core (inner cavities and a potential outer cavity in the E), a strong radio point source in the center of the BCG, and H$\alpha$ filaments. These features are either absent or less pronounced in the northern BCG, which we describe in more detail in section \ref{ch:north_core}. 
\cite{OSullivan2019-le} did not find extended radio emission in a GMRT pointed observation at $\SI{1.39}{GHz}$ in the southern BCG of NGC~6338, at least not beyond a small extension of the core to the south-west. However, assuming a typical spectral index for aged radio lobes we expect our new uGMRT datasets to be at least 5 times deeper, and able to detect any signs of past feedback. 

\subsubsection{Extended radio emission}
\begin{figure*}
    \centering
    \includegraphics[width=0.99\textwidth]{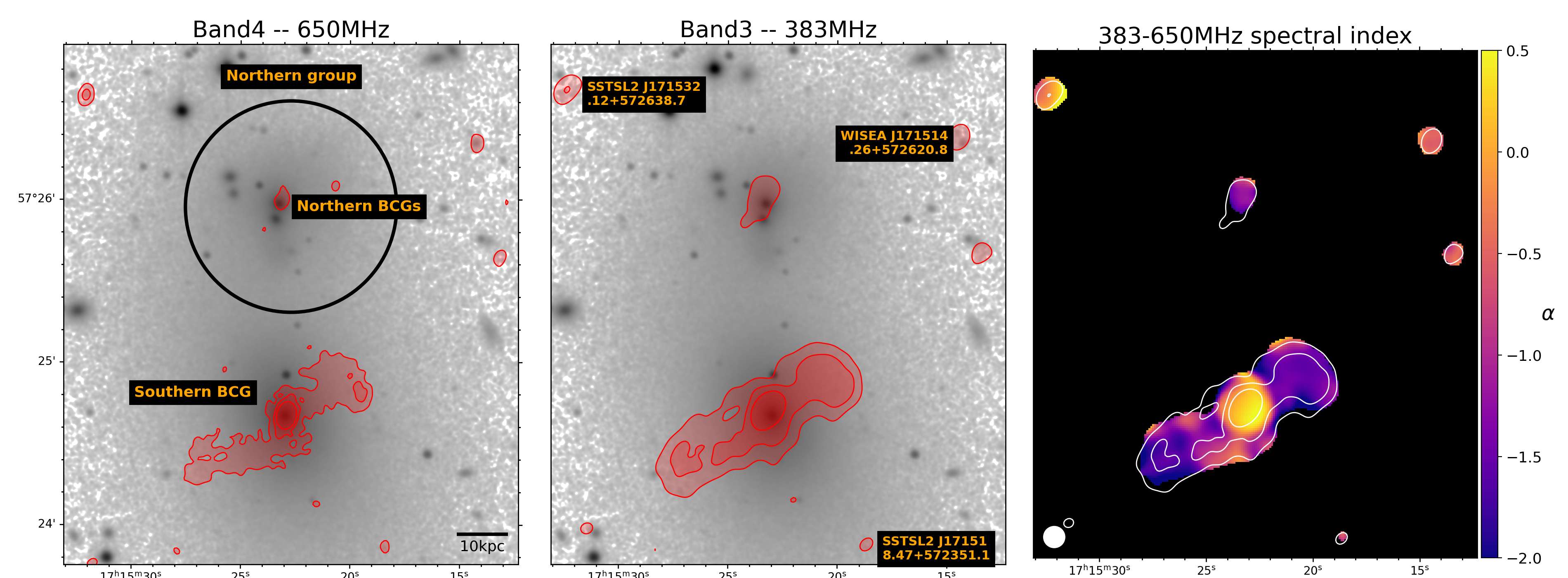}
    \caption{Radio emission in the center of the NGC~6338 group: \textit{Left:} SDSS r-band image with uGMRT Band 4 contours at  5, 25, 125 $\sigma$ levels. \textit{Middle:} As left panel but for uGMRT Band 3. \textit{Right:} Spectral index map of uGMRT band 3 and 4 data. The white contours show the emission at 383\,MHz at 5, 25, 125 $\sigma$. The restoring beam is shown in the lower left corner.}
    \label{fig:radio}
\end{figure*}
Low frequency $\SI{143}{MHz}$ LOFAR data has detected extended emission around the southern core of NGC~6338 (\citealp{Birzan2020-oe}, and Fig. \ref{fig:lofar}). The restoring beam size of LOFAR is $13.7\times 8.7\,\si{arcsec}$, which is about 3 times larger than the uGMRT band 4 data (Tab. \ref{tab:gmrt}), while uGMRT also has sufficient sensitivity to detect the extended emission. 
Figure \ref{fig:radio} shows the radio contours uGMRT band 3 (middle panel), and uGMRT band 4 (left panel) on an SDSS r-band image. The band 3 and 4 data clearly resolve the structure of this extended emission: We see a pair of lobes in the south-east north-west direction, which have not been detected at 1.4\,GHz with the GMRT (\citealp{OSullivan2019-le}). 
The uv-coverage of the GMRT data is sensitive to extended emission on scales up to $\SI{3.4}{arcmin}$ (band 4), and $\SI{4.3}{arcmin}$ (band 3). 
The south-east lobe extends about $\SI{20}{kpc}$, while the north-west lobe extends out to only 14\,kpc. 
Both lobes are detected at high significance (contours in Fig. \ref{fig:radio} start at 5$\sigma$). 
The flux density in the south-east lobe is $\SI{10.0(1)}{mJy}$ at band 3 and $\SI{5.1(1)}{mJy}$ at band 4, while the other lobe has $\SI{11.4(1)}{mJy}$ and $\SI{6.1(1)}{mJy}$ at band 3 and 4, respectively. The peak emission in the southern BCG coincides with the optical center of the galaxy and is $\SI{56.2(1)}{mJy/bm}$ in band 3, and $\SI{62.2(1)}{mJy/bm}$ in  band 4, yielding a slightly positive spectral index. Since the flux density at this location might be blended with the lobes it is only an upper limit for the flux of the central AGN.
Next to the center of the AGN we can identify another, much smaller pair of lobes in the orthogonal direction (south-west to north-east), which we describe in more detail below. 

We find some other (distant) radio sources in the nearby field that we can identify using NED (see Fig. \ref{fig:radio} middle panel) and confirm the accuracy of our astrometric corrections. We do not find other extended sources in the center of this merging group, and none in the outskirts that are related to NGC~6338. The brightest source in the field ($\SI{136}{mJy\,bm^{-1}}$ in band 3) is about $\SI{20}{\arcmin}$ to the west.

The depth of the uGMRT data in both bands allows us to create a spectral index map from 383 to 650\,MHz for the southern BCG. Figure \ref{fig:radio} (right) shows a map of the spectral index $\alpha$, which we define as $S_\nu \propto \nu^\alpha$, where $S_\nu$ is the flux density at frequency $\nu$. 
We show the associated error map of the spectral index in Fig. \ref{fig:spixerr}.
The core of the region exhibits an inverted spectrum with $\alpha_{\rm 383-650\,MHz}=\num{0.34(5)}$. The spectrum steepens when moving away from the core, and reaches $\num{-1.64(8)}$ in the south-east, and $\num{-1.49(5)}$ in the north-west. 

As shown by the band 3 contours in Fig. \ref{fig:radio} (right), the steepening does not follow exactly along the trajectory given by the brightness distribution. 

\begin{figure}
    \centering
    \includegraphics[width=0.45\textwidth]{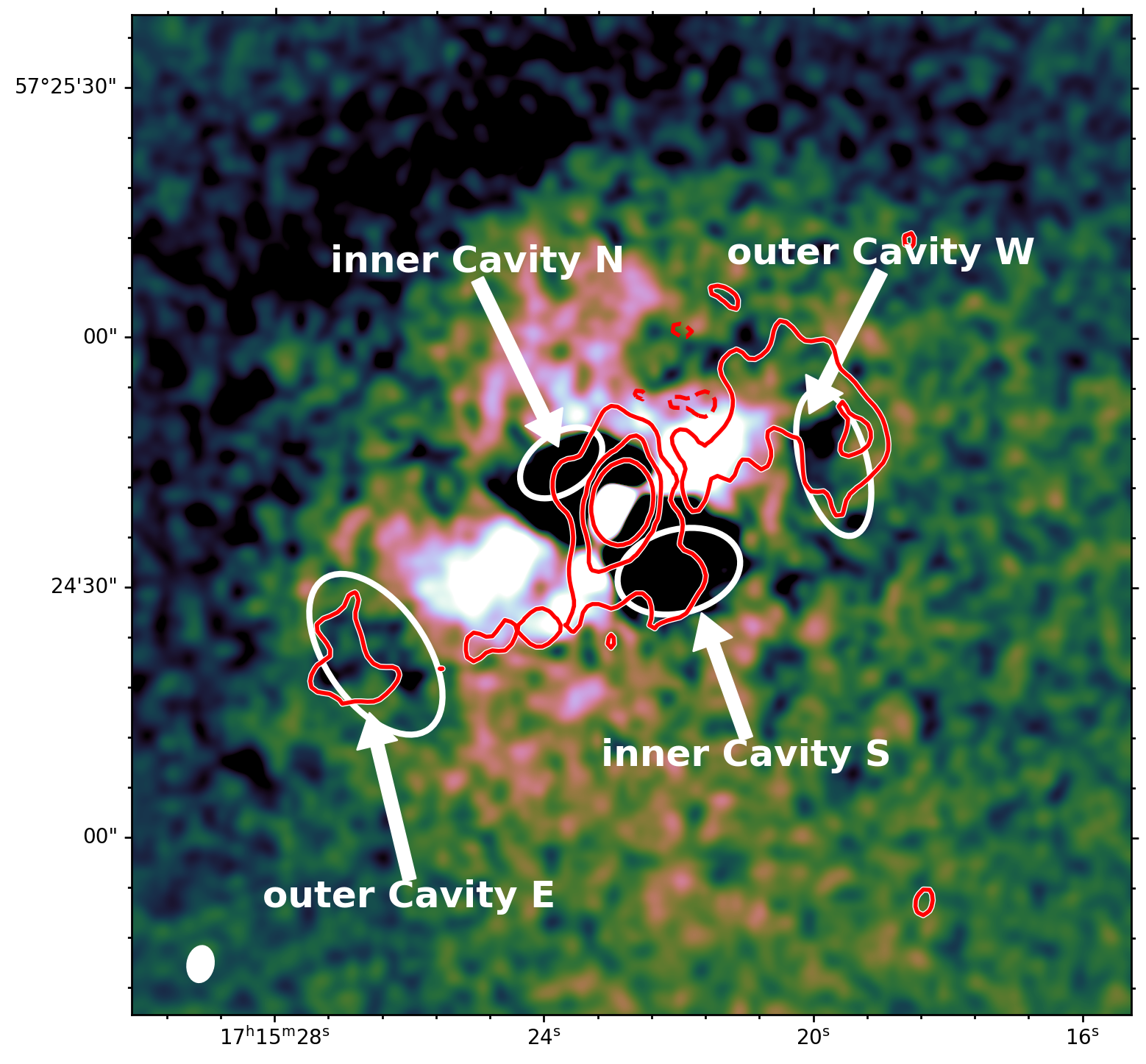}
    \caption{Chandra X-ray residual map with high resolution (excluding short baselines, see text for details) uGMRT Band 4 contours (red, 5, 25, 125 $\sigma$, negative 5 $\sigma$ shown as dashed contours). Note that most of the large-scale radio emission is resolved out to highlight the emission in the inner cavities. The beam is shown in the lower left corner.}
    \label{fig:cavities}
\end{figure}
\begin{figure}
    \centering
    \includegraphics[width=0.45\textwidth]{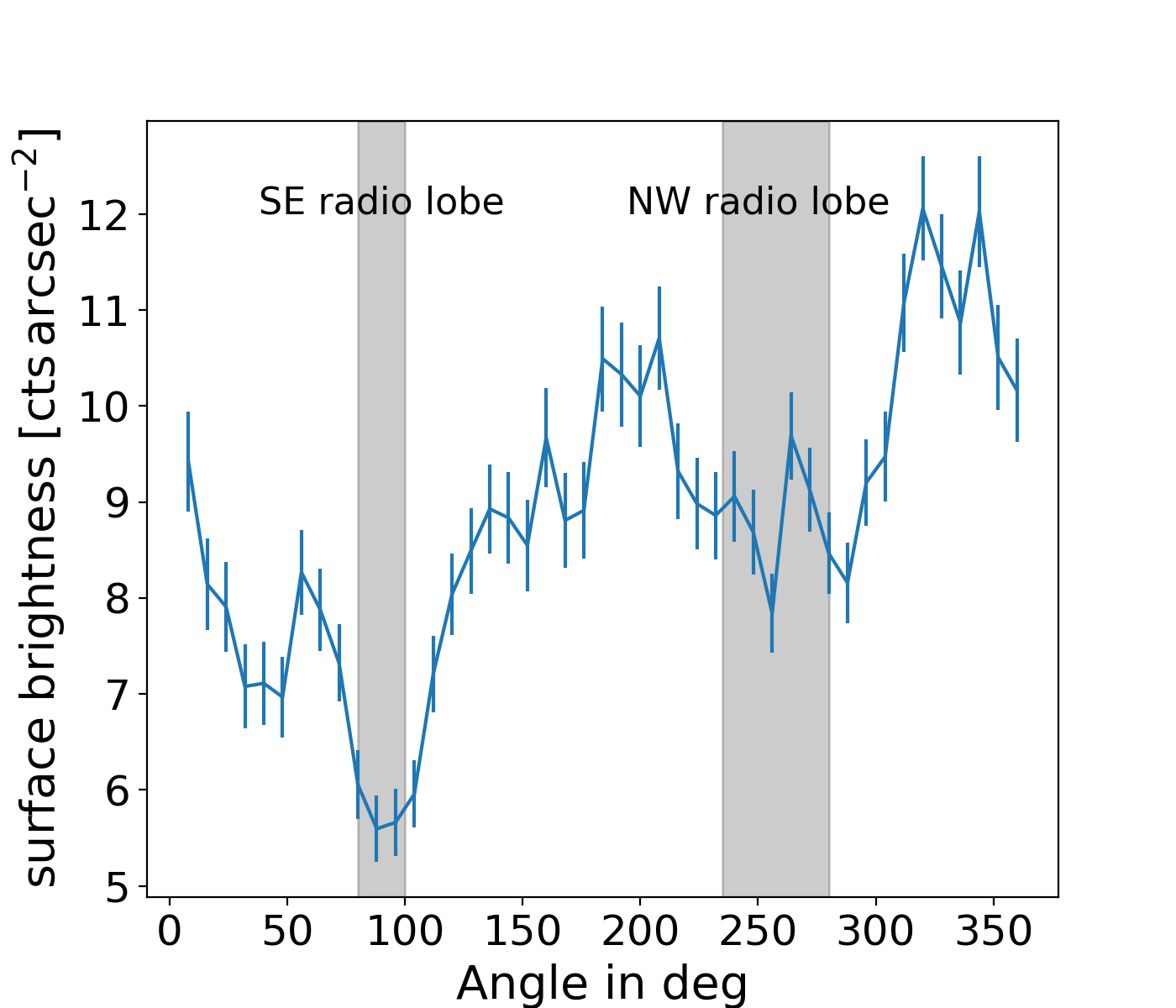}
    \caption{Azimuthal profile from the Chandra $\SIrange{0.5}{2}{keV}$ image at the location of the outer radio lobes.}
    \label{fig:azimuth}
\end{figure}

The radio lobes in the southern BCG described above mark an older outburst of the BCG. It is expected that these radio-bright relativistic particles evacuate a confined region of the hot gas, which will move to larger radii due to buoyant forces. We tested if the region containing the radio plasma actually coincides with cavities in the X-ray image by overlaying it with the Chandra residual map (from \citealp{OSullivan2019-le} Fig. 3 left). 
\cite{OSullivan2019-le} found a cavity to the SE, and from their X-ray residual map a less significant depression at a similar distance to the NW can be identified (see outer cavities in Fig. \ref{fig:cavities}). 
We created an azimuthal profile from the Chandra $\SIrange{0.5}{2}{keV}$ image in an annulus at the distance of the radio lobes (Fig. \ref{fig:azimuth}). We confirm the eastern cavity (Fig. \ref{fig:cavities}) that has been tentatively detected by \cite{OSullivan2019-le}. The western cavity shows a dip in the azimuthal profile, but the surface brightness depression is much broader than the eastern cavity. 
Unlike the eastern cavity, where a rim structure is clearly visible, we cannot find this around the western cavity. 
Although the radio lobe location matches the surface brightness dip of the western cavity, we can only classify this as a possible X-ray cavity, also due to the complex structure of the whole system.

The eastern cavity and western cavity candidate coincide with the location of the old radio lobes (SE and NW, labelled as outer cavities). The red contours in Fig. \ref{fig:cavities} are derived from a band 4 image that excludes part of the large-scale emission by applying a minimum uv-threshold of 8k$\lambda$, which not only points out the inner lobes mentioned earlier, but also shows only the brighter patches of the outer cavities. The inner lobes are perpendicular to the outer lobes (in projection), and seem to largely fill the strong X-ray inner cavities. 

\subsubsection{Radio spectrum}\label{ch:radio_spectrum}
\begin{figure*}
    \centering
    \includegraphics[width=0.49\textwidth]{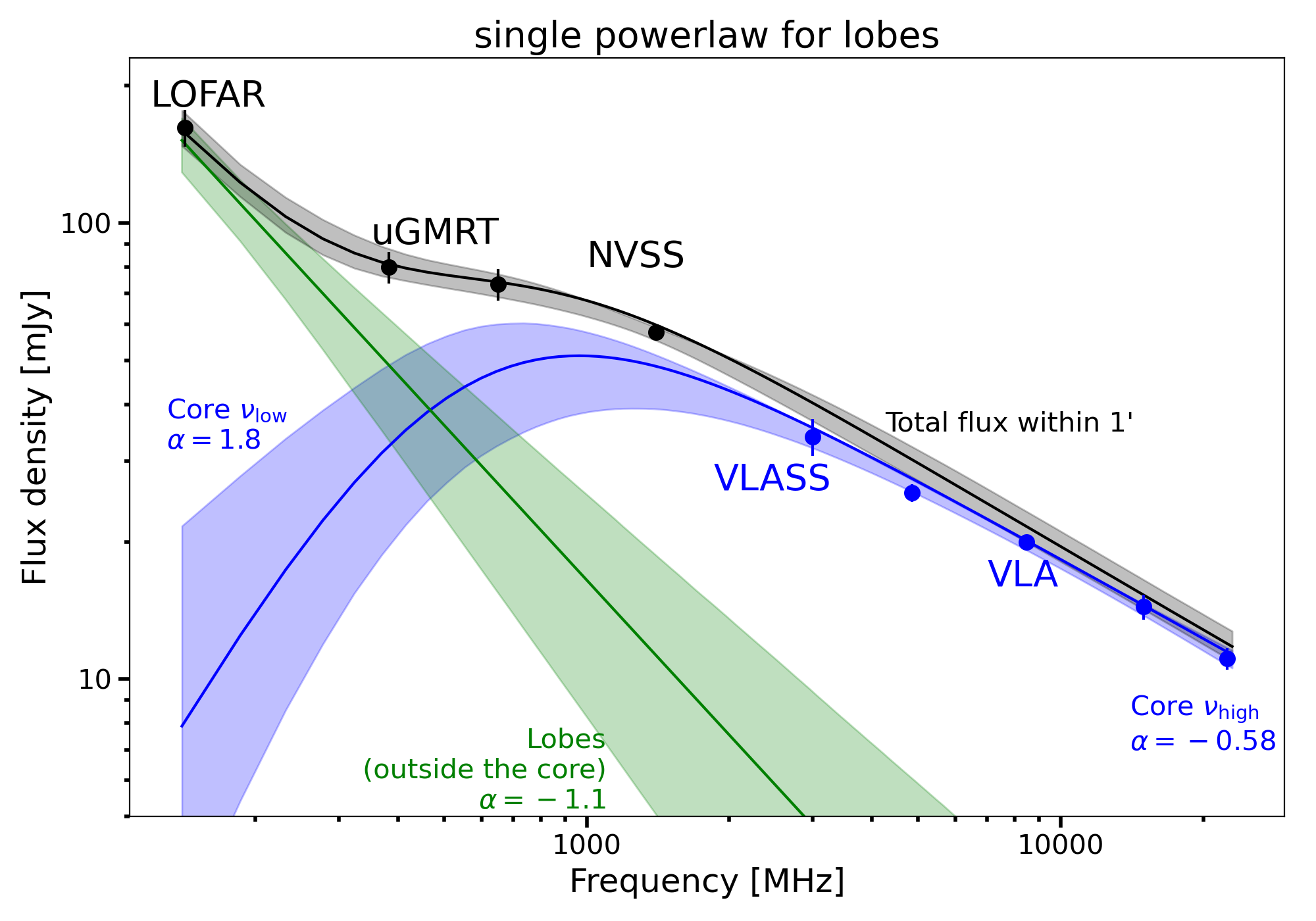}
    \includegraphics[width=0.49\textwidth]{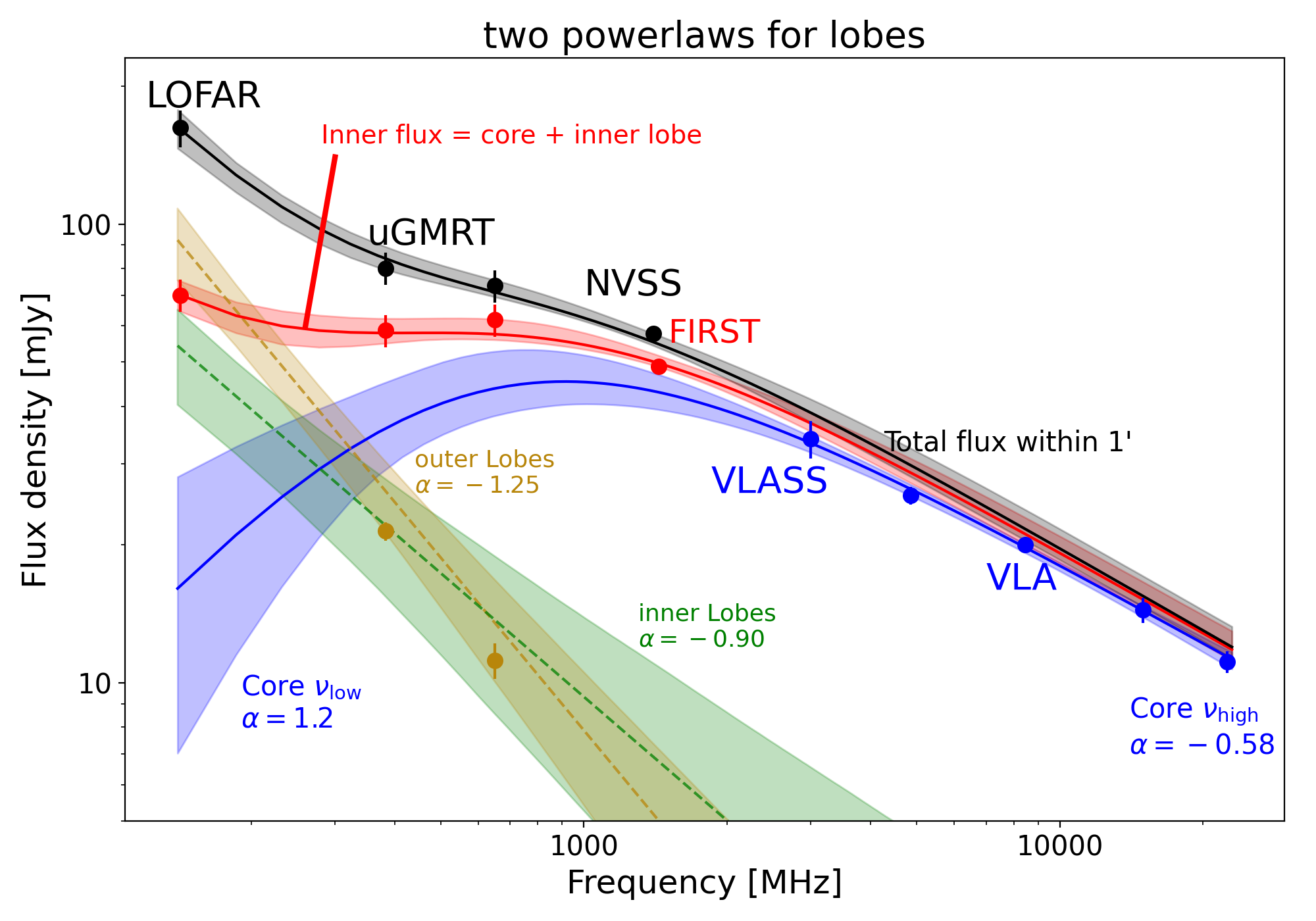}
    \caption{Radio spectrum of the southern core. Black datapoints and black model lines show the total flux enclosing a large region that includes the core and the lobes, while the blue datapoints and model lines show only the core flux. \textit{Left} for the case of a single powerlaw (green) for the lobes, and \textit{right} for the case of two powerlaws for the inner (green dashed) and outer lobes (yellow dashed). In the case of two powerlaws, we also fit measurements from a inner region that encloses the core and inner lobes (red is blue plus dark red). }
    \label{fig:spectrum}
\end{figure*}
In the previous paragraph we have described the spatial distribution of the extended radio emission, especially in the southern BCG. With our multi-frequency coverage, we are able to also analyze the integrated radio spectrum. Therefore, we include more datasets, beyond the two uGMRT bands: 
We have a low frequency measurement from LOFAR at $\SI{142}{MHz}$, which shows extended emission, but does not allow us to cleanly separate it from the central core. We also have several measurements above 1\,GHz: 
\begin{itemize}
    \item The NRAO VLA Sky Survey at $\SI{1.4}{GHz}$ (NVSS, \citealp{Condon1998-tu}) with $\SI{45}{\arcsec}$ resolution is sufficient to measure the flux of the integrated emission, but cannot resolve any structure. 
    \item The VLA FIRST Survey at $\SI{1.44}{GHz}$ (\citealp{Becker1994-te}) provides a spatial resolution of $\SI{5}{\arcsec}$. It allows us to measure the core emission blended with the inner lobes, and is sensitive to detect extended emission on scales up to 2\,arcmin.
    \item The Karl G. Jansky Very Large Array Sky Survey (VLASS, \citealp{Lacy2020-kg}) at $\SI{3}{GHz}$ is a currently ongoing survey, which has already completed two epochs with scans of the NGC~6338 field. Its $\SI{2.5}{\arcsec}$ spatial resolution allows a clear detection and measurement of the core flux. No extended emission is detected, while it should be able to detect emission on scales up to $\SI{1}{arcmin}$, which is almost the size of the outer lobes.. 
    \item The VLA has observed the NGC~6338 field at a number of frequencies since 1990, which allows us to track the time variability of the AGN emission. This will be discussed in Section \ref{ch:timevar}. Here we describe the spectrum up to 25\,GHz and make use of the simultaneous observation in C, X, Ku, and K band ($\SI{4.86}{GHz}$, $\SI{8.46}{GHz}$, $\SI{14.94}{GHz}$, and $\SI{22.46}{GHz}$, respectively) in project AM701 (Tab. \ref{tab:vla}). Each band allows a clear flux measurement, but no band shows signs of extended emission or any deviation from a single point source. 
\end{itemize}

Our modeling of the SED consists of a core emission model and a model for the lobe emission. 
A simple model that parameterizes the core emission is a powerlaw with a soft absorption part to account for decreasing fluxes at lower frequencies. 
The physical interpretation of the absorption model is discussed in section \ref{ch:discussion_agn}. 
In our model for the core emission, we leave the spectral index at low frequencies free to vary, and also include a variable turn-over frequency and a smoothing parameter. The model for the lobe emission is a simple powerlaw. 

The fluxes from LOFAR, uGMRT, and NVSS are extracted within an aperture of $\SI{1}{arcmin}$ radius, which includes all the emission from the outer lobes and the core, but not any emission from the northern BCG. We show these ``total'' fluxes as black datapoints in Fig. \ref{fig:spectrum}, to which the model of the core plus the lobe is fitted. 
We note that the shortest LOFAR baselines are 90m at $\SI{140}{MHz}$, corresponding to a largest angular scale (LAS) of $\SI{1.3}{\deg}$. However, the LOFAR image does not show any emission on scales larger than $\SI{1.5}{arcmin}$. The uGMRT datasets are imaged from $700\,\lambda$, which corresponds to an LAS $\SI{4.9}{arcmin}$. Therefore, we can safely compare the fluxes measured from LOFAR and uGMRT. However, for the spectral index maps (Fig. \ref{fig:radio} right) all datasets were imaged with identical uv ranges.
The measurements above 2\,GHz (shown in blue in Fig. \ref{fig:spectrum}) detect only the core emission, and therefore only the core model is fitted in this case.
The fitting is done with the emcee MCMC framework (\citealp{Foreman-Mackey2013-hw}) by modelling the likelihood for the two components in a single function. All parameters have flat priors.  

The fit with this simple powerlaw for the lobe emission is shown in the left panel of Fig. \ref{fig:spectrum}. We find  that the resulting model fits the data well ($\chi_{\rm red}^2 = 2.5$, ${\rm dof}=2$). The low frequency index of the core emission is relatively close to the theoretical 2.5, but not above, making synchrotron self-absorption (SSA) a valid possibility. The high frequency index of $-0.58^{+0.05}_{-0.07}$ is typical for AGNs with active jets. The lobe model with a spectral index of $-1.13^{+0.23}_{-0.33}$ contributes most of the flux to the LOFAR measurement, and about half the flux of the integrated source at uGMRT band 3.

The spatial resolution of the uGMRT data allow us to distinguish the lobe emission from the inner region flux (core plus inner lobes). 
The lobe flux at $\SI{383}{MHz}$ is about $\SI{21}{mJy}$, while the model  for the single powerlaw (green line in Fig. \ref{fig:spectrum}) predicts about $\SI{47}{mJy}$. The flux of the inner core region at the GMRT frequency is also much higher than the simple core model (blue line in Fig. \ref{fig:spectrum}).  
This means we are unable to link the derived model components to physical regions. Therefore, we refine the lobe model by adding a second powerlaw component, so that one powerlaw accounts for each set of lobes (inner and outer). 
We included the flux measurements of the inner region  including both the core and the inner lobes, but we do not spatially separate the inner lobes from the core since the overlap is too large and projection effects would be inevitable. The resulting fit is shown in Fig. \ref{fig:spectrum} right: The integrated flux (black) and the core flux (blue) are as well fit as for the simple case. The model for the outer lobe (yellow) agrees with measurements of the outer lobe flux at the uGMRT frequencies, although these measurements have not been included in the fit (yellow datapoints). The outer lobe model shows a steeper spectrum than the inner lobes ($-1.25^{+0.20}_{-0.22}$ vs. $-0.90^{+0.23}_{-0.26}$), indicating an older plasma. This is in agreement with the impression from the spectral index map in Fig. \ref{fig:radio}.
The other model components change only slightly when including the second powerlaw for the lobes: 
The core high frequency spectral index remains at $-0.58^{+0.05}_{-0.08}$, and the low frequency spectral index changes only marginally to $1.2^{+1.2}_{-0.5}$. We also find that the turnover frequency for the core model changes slightly within the uncertainties (from $923^{+371}_{-191}\,\si{MHz}$ to $849^{+243}_{-168}\,\si{MHz}$). The overall fit improves ($\chi_{\rm red}^2 = 1.1$, ${\rm dof}=4$). 

\subsection{Did the AGN in the southern BCG change its direction of feedback?}
The new, inner lobes that coincide with the inner X-ray cavities (Fig. \ref{fig:cavities}) indicate a change of direction of the feedback provided by the AGN, since they are not aligned with the tentative outer cavities. The change appears to be almost $90^\circ$ in projection, which is significant. 
\begin{figure}
    \centering
    \includegraphics[width=0.45\textwidth]{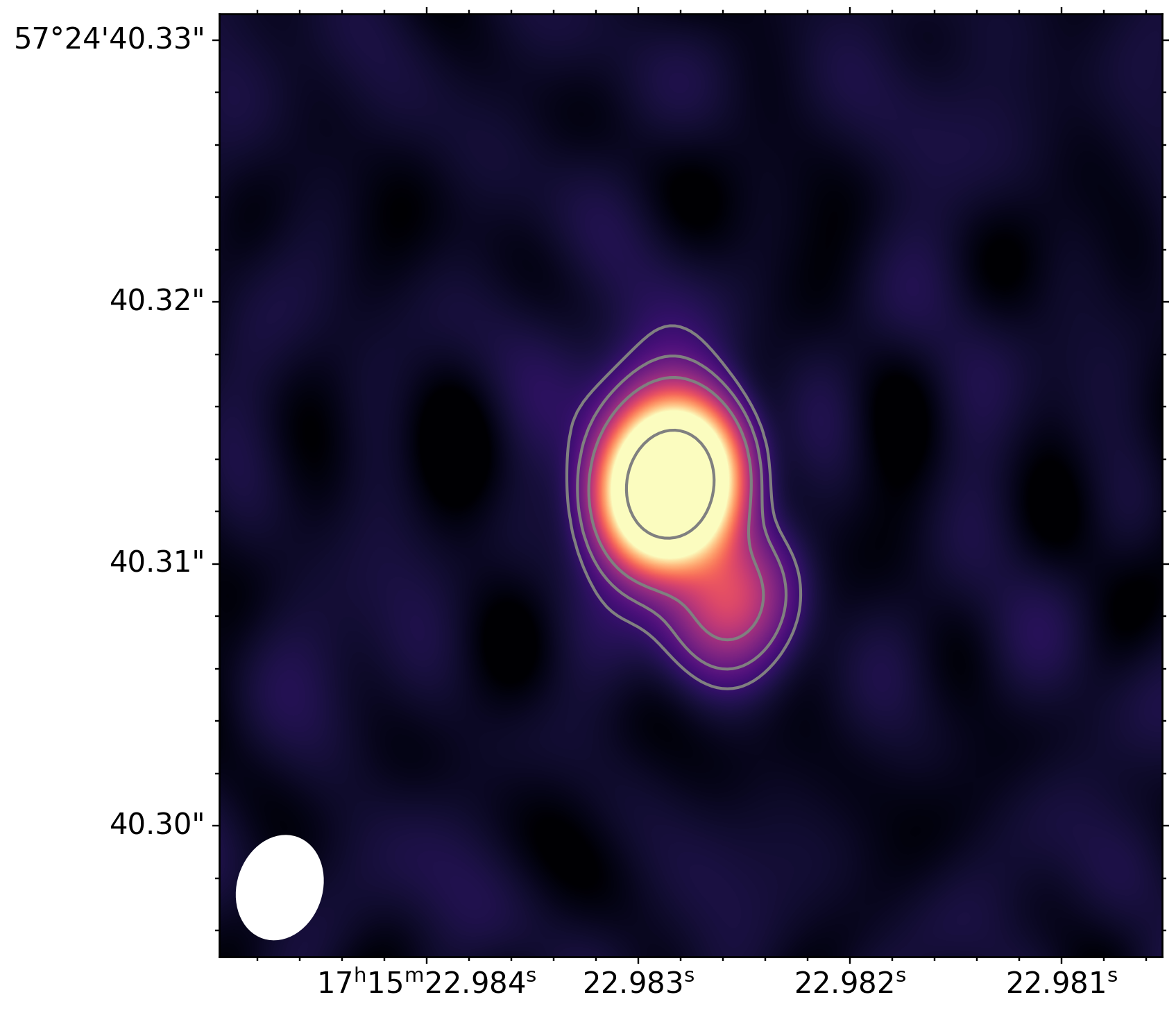}
    \caption{VLBA image of the southern core with contours at the 5, 10, 20, and 100$\sigma$ levels. The restoring beam is shown in the lower left corner. }
    \label{fig:vlba}
\end{figure}
To understand the current state of the AGN, in particular if the AGN is currently active, we analyzed the archival VLBA observation that targeted the center of the southern BCG. Our VLBA image (Fig. \ref{fig:vlba}) shows a bright source in the center, possibly the core, with R.A. $17:15:22.98$ and Decl. $+57:24:40.31$. 
A fainter point source is located $\SI{5.1}{mas}$ (about $\SI{2.7}{pc}$) to the south-west. 
Since we do not detect a double sided jet with a core in the center, we cannot say for sure, which of the two sources is the core, since hotspots can exceed the core in flux (e.g., \citealp{Taylor1996-sp}). However, we can infer an orientation angle from the location of the two sources, which is close to $45^\circ$, and consistent with the orientation of the inner lobes. 
We use the CASA task \textit{imfit} to fit two Gaussian components to the brightness distribution. This results in $\SI{21.5(2)}{mJy}$ for the brightest source, and $\SI{3.4(3)}{mJy}$ for the source offset to the SW. 

The total VLBA detected flux density at $\SI{4.98}{GHz}$ is $\SI{25.7(3)}{mJy}$. The deepest C-band VLA observation is also very close in time (within a year), and finds a flux of $\SI{28.1(1)}{mJy}$ at the effective frequency $\SI{6}{GHz}$. Utilizing the spectral index found in the previous section, this converts to a flux density of $\SI{31.0(1)}{mJy}$ at the VLBA frequency. This 20\% difference in flux between VLBA and VLA could be due to short-term flux variability. We also note that another C-band observation in 2001 obtained the same flux density as the VLBA. 
But even if 20\% of the flux is missed by the VLBA, it means that most of the core emission is on scales smaller than $\SI{43}{mas}$ ($\SI{22}{pc}$), which marks the angular size of the shortest VLBA baseline. 
We are confident that we are not missing other jet emission in the direction, e.g., of the older lobes, and that  the change in jet direction is real.

\subsection{What is the current state of feedback in the southern BCG?}
\subsubsection{High frequency flux}
\begin{figure}
    \centering
    \includegraphics[width=0.5\textwidth]{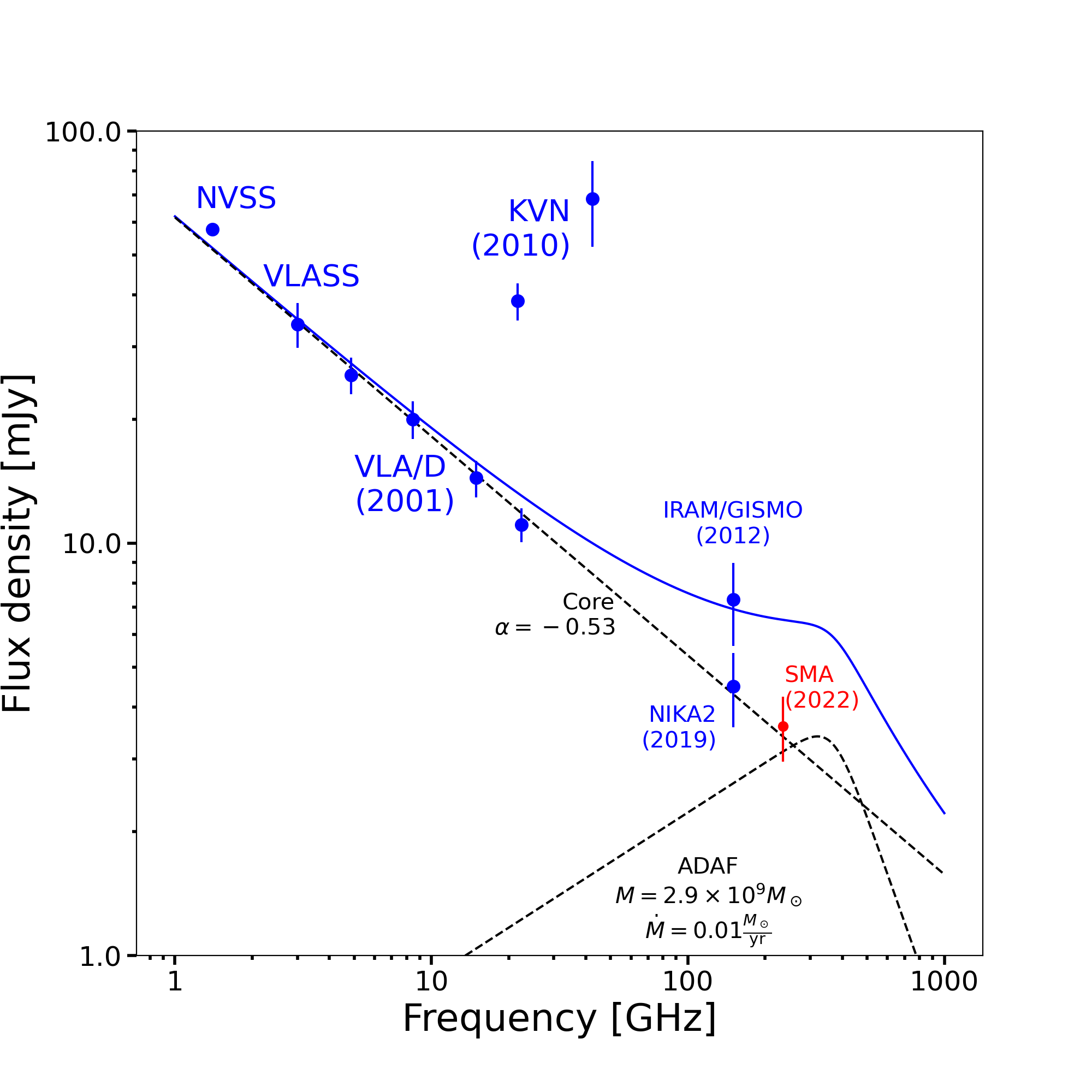}
    \caption{Modeling of the high frequency spectrum of the southern BCG. The dashed powerlaw is consistent with all datapoints except the IRAM/GISMO measurement, which requires an additional ADAF component, that does not match the current SMA flux. KVN measurements use a large aperture and might be biased.}
    \label{fig:sma_flux}
\end{figure}
The VLBA image in Fig. \ref{fig:vlba} suggests a small jet in the northeast-southwest direction. However, the observation is not deep enough to constrain a separate spectral index for the core and jet component, which would be favorable for age estimates. 
Therefore, we extend the radio spectral analysis presented in Fig. \ref{fig:spectrum} to even higher frequencies up to 230\,GHz. This allows us to test for an accretion disk model component, such as an advection dominated accretion flow (ADAF, e.g., \citealp{Mahadevan1997-rx,Narayan1998-be}). For example Sgr A$^\star$ is far below Eddington luminosity and the SED from radio to X-rays can be model with an ADAF component for the accretion as demonstrated by \citealp{Yuan2002-if}. 
\cite{Schellenberger2020-ji} confirmed that the ADAF model could account for the high frequency flux increase in NGC5044, and interpreted this as an active feedback cycle. 

The Korean VLBI Network facility comprises three 21m radio telescopes with baselines from 300 to 500\,km. \cite{Park2013-lv} observed NGC~6338 simultaneously at 22 and 42\,GHz with two antennas in single-dish mode on Dec 31, 2010. The fluxes (Fig. \ref{fig:sma_flux}) are much higher (more than a factor of 3) than what was expected from the VLA measurements and the powerlaw model. However, \cite{Park2013-lv} note that for these two measurements the pointing correction could not be done since the source was only detected in either the azimuth or elevation cross-scan, but not in both. The source fluxes are therefore unreliable, and it remains unclear if the inconsistent fluxes are due to false measurements, or time variability. Therefore, we do not take these two datapoints into account for a spectral fit.

NGC~6338 has also been observed with the GISMO bolometer camera on the IRAM-30m telescope in Oct 2012 by \cite{Hogan2015-uv}. The authors showed NGC~6338 to have a flux of $\SI{7.3(15)}{GHz}$ at 150\,GHz. The authors interpreted the GISMO flux as ``flickering  at  high  frequencies  or  variability''. It is also slightly inconsistent with the re-observation in 2019 by \cite{Rose2021-wy} with the NIKA2 bolometer on the IRAM-30m telescope, which found the flux to be $\SI{4.5(5)}{mJy}$ at the same frequency as GISMO. Note that we averaged here the two NIKA2 two measurements which were about 7 months apart, and are consistent with each other. 

In the ADAF model the flux density is proportional to the accretion rate $\propto \dot m^{4/5}$ for the synchrotron part of the spectrum (\citealp{Mahadevan1997-rx}).
If we interpret the GISMO measurement as indications for ongoing accretion and model it using an ADAF model (see Fig. \ref{fig:sma_flux}), we find a small accretion rate of $\SI{0.01}{M_\odot\,yr^{-1}}$ for a black hole mass of $\SI{2.9e9}{M_\odot}$ (as determined by \citealp{Mezcua2018-ml} from the K-band). 
This accretion model allows us to model the GISMO flux, but is also inconsistent with NIKA2. 

Most recently, our dedicated SMA observation at 230\,GHz revealed the current flux to be consistent with the NIKA2 measurement, assuming the powerlaw spectral index of $-0.53$ (see Fig. \ref{fig:sma_flux}). If the GISMO, NIKA2 and SMA measurement are all taken at face value, one could imagine the accretion rate only affecting the time around 2012 (GISMO measurement), and decreasing to lower values afterwards. 

\subsubsection{Time variability}
\label{ch:timevar}
\begin{figure}
    \centering
    \includegraphics[width=0.5\textwidth]{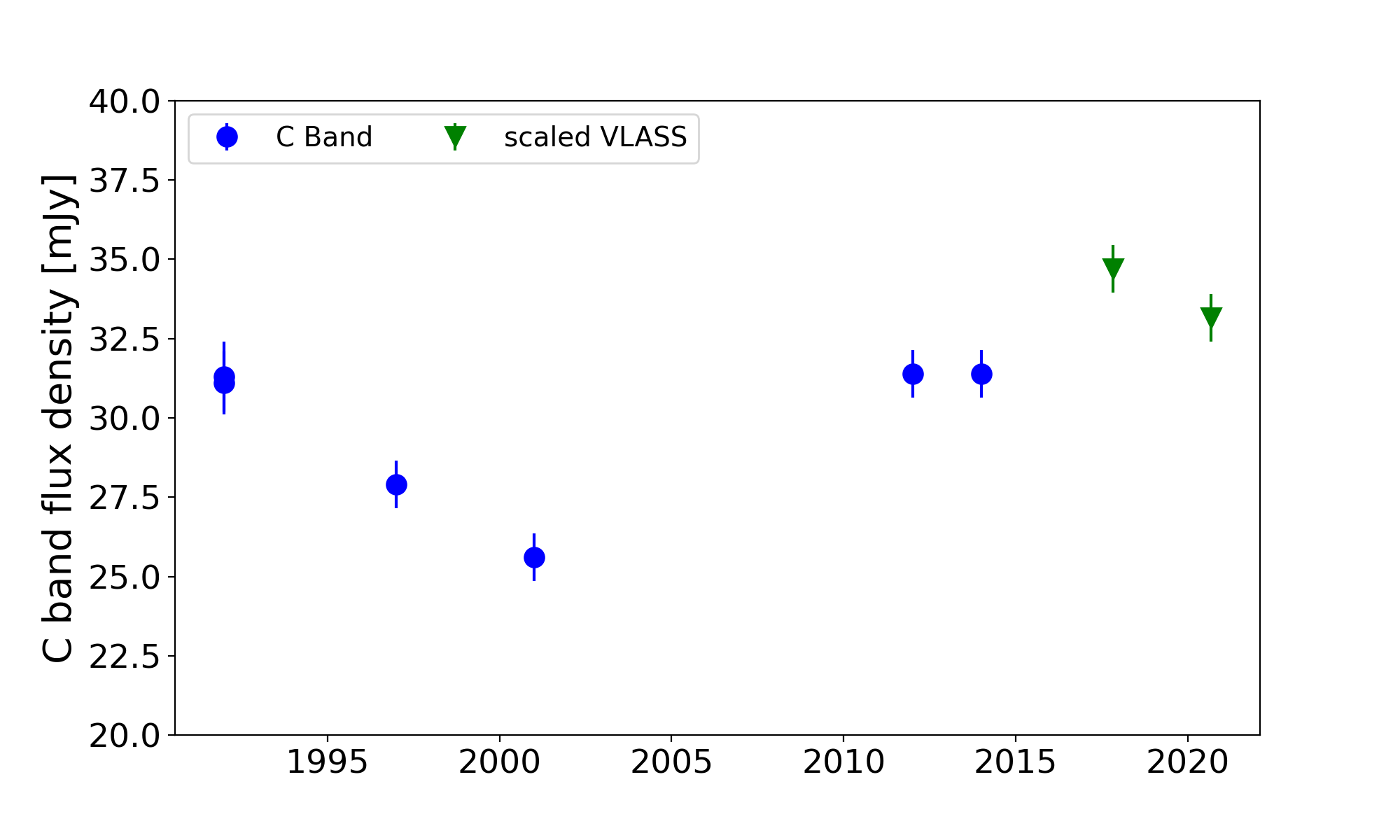}
    \caption{Lightcurve of the southern core flux of VLA observations in S band (2.3\,GHz), and C band (5\,GHz) normalized to 5\,GHz using a spectral index of $-0.58$. }
    \label{fig:lightcurve}
\end{figure}
A time variability of the accretion flow might be linked to the size of the accretion disk.
Unfortunately, we have no sufficient time sampling of the lightcurve at mm wavelengths available for NGC~6338, and the time variability study relies on the frequent VLA observations. 
The VLA frequencies between 1 and 5\,GHz won't be affect much by a moderate ADAF component (see Fig. \ref{fig:sma_flux}). 
However, variability in these bands can be connected to jet disturbance or in-jet shocks from a largely turbulent magnetic field (e.g., \citealp{Aller2017-di,ODea2021-rn}). 
Core-dominated flat-spectrum radio AGN typically show the most dramatic examples of intensity and polarization variability. 
We have analyzed repeated VLA observations (Tab. \ref{tab:vla}) of the AGN in the southern BCG, which allows us to construct a lightcurve (Fig. \ref{fig:lightcurve}). 
Over the past 30 years, NGC~6338 has been observed 6 times in C band (blue points in Fig. \ref{fig:lightcurve}), and twice in S band for the VLASS survey (green points in Fig. \ref{fig:lightcurve}). We treat the C band flux densities as a reference since they provide the best long term coverage, and convert VLASS fluxes densities to 5\,GHz using the spectral index $-0.53$. We decided to exclude the L band fluxes here because of the changing contribution from extended emission (inner and maybe outer lobes) in different VLA configurations. The extended emission has only minimal contribution to the flux density at C band in the most compact configuration ($\lesssim \SI{0.7}{mJy}$). The only C band observation in D configuration was taken in 2001, where we find the lowest flux. Subtracting any extended emission will result in an even lower point source flux and increase the variability. 
The overall scatter (standard deviation, not taking into account the time of observation) between the C band observations is $\SI{3.8}{mJy}$, exceeding by far the typical statistical uncertainty (average r.m.s) of $\SI{0.2}{mJy}$. The same is true for the L band observation, where the scatter is $\SI{5.4}{mJy}$ and the average r.m.s about $\SI{1}{mJy}$. 
The smoothed lightcurve over all fluxes shown as black line in Fig. \ref{fig:lightcurve} shows a decrease in flux until 2005, and a rise afterwards. The region around this turnover is not sampled well. The magenta line in Fig. \ref{fig:lightcurve} shows the relative change in flux per year based on the smoothed curve, which also shows the constant 2\% decrease until 2005, and the peak of 6\% increase shortly after that. The typically variability is on timescale of years, but we lack good short term sampling. 

\subsection{Did the feedback stop in the northern BCG?}\label{ch:north_core}
\begin{figure}
    \centering
    \includegraphics[width=0.44\textwidth]{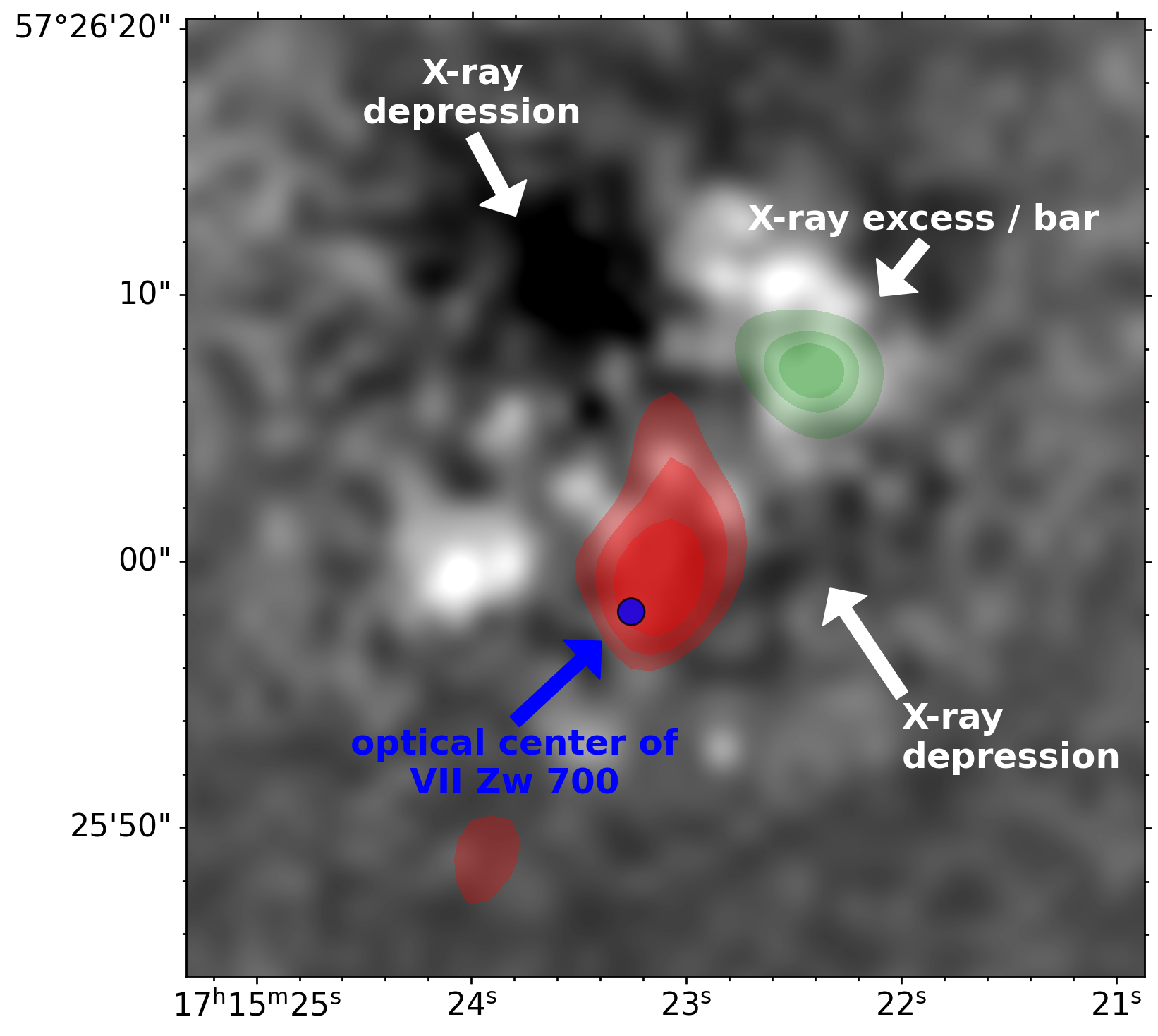}
    \caption{X-ray residual image of the northern core. Green filled contours show the H$\alpha$ (\citealp{OSullivan2019-le}) emission, red contours the 650\,MHz uGMRT data, and the blue circle shows the optical center of the galaxy.}
    \label{fig:north_core}
\end{figure}
The merging group toward the north appears to host two bright galaxies in the center (see Fig. \ref{fig:radio}). 
We detect radio emission associated with the northern of the two in both uGMRT bands with high significance, $\SI{1.4(1)}{mJy}$ at band 3, and $\SI{0.59(10)}{mJy}$ at band 4. This indicates a very steep spectrum source with $\alpha_{383-650\,{\rm MHz}}=-1.6^{+0.3}_{-0.4}$. Since the LOFAR flux at this location is only $\SI{3.2(3)}{mJy}$, the spectrum of the northern source is likely bent, e.g., due to aging of radio lobes that are not resolved here. The spectral index between LOFAR and uGMRT band 3 is significantly flatter $\alpha_{142-383\,{\rm MHz}}=-0.83^{+0.12}_{-0.12}$. No emission is detected in any VLA observation (L, S, or C band), but the noise levels do not allow to claim an upper limit for an even steeper spectral index toward these high frequencies. 

\cite{OSullivan2019-le} found several features in the X-ray residual map, such as a emission aligned in a southeast-northwest oriented bar, with depressions in the X-ray on either side of the bar. The depression in the southwest appears to be enclosed by a brighter rim, marking it a likely cavity.
However, we find that the faint radio emission is not aligned with the cavities, but seems to extend from the center of the bar toward the south-south-east (see Fig. \ref{fig:north_core}). The X-ray center, the brightest part of the southeast-northwest bar, is offset to the north of the optical center of the galaxy (blue mark in Fig. \ref{fig:north_core}). 
The uGMRT radio emission in the northern core has a head-tail structure in Band 4, with the tail toward the north, and the head coinciding with the optical center. Some fainter emission is offset to the south-east, and the Band 3 data covers this blob as well. 
The projected image in Fig. \ref{fig:north_core} indicates that the H$\alpha$ emitting gas and the radio plasma share a similar direction.  \cite{Lin2017-wa,Pan2020-ei} show IFU data that the H$\alpha$ filaments extend from the optical center to the green region in Fig \ref{fig:north_core}. The location of the warm gas and the lobes from the AGN give a consistent picture, while the X-ray cavities are offset from this. However, since the merger is mainly along the line of sight, there are strong projection effects to be considered here: The main extent of the radio tail is likely not to the north, but along the line of sight.

\section{Discussion}
\label{ch:discussion}
NGC~6338 is a violent merger with two visible cores. The merger is mostly along the line of sight with an offset of a few hundred kpc. We have examined the visible signs of feedback in the southern BCG: We have detected a set of old lobes in our deep uGMRT observations at 383\,MHz and 650\,MHz, which extend 45\,arcsec (24\,kpc) in the south-est and north-west direction. The radio lobes match the location of X-ray cavities in the IGM. 

\subsection{Age of the past feedback cycle}
We observe strong features of AGN feedback in the southern BCG of the merging galaxy group NGC~6338, such as X-ray cavities filled with radio lobes, a smaller set of lobes, and a radio bright core that seems to have a parsec-scale jet in the VLBA image. However, no CO was detected, placing an upper limit on the amount of cold gas of $\SI{7.3e7}{M_\odot}$, which is plausible for a star formation rate on the order of $\SI{0.1}{M_\odot\,yr^{-1}}$ (\citealp{Crawford1999-cw,OSullivan2018-yk}). 
NGC~5044, which resembles a typical relaxed, cool core galaxy, has been found to have a comparable amount of cold gas from a cooling hot gas phase (\citealp{Schellenberger2020-vl}). Therefore, the derived upper limit for NGC~6338 is in-line with expectations for a typical cool-core galaxy group. 

With the available information from our radio analysis we are able to quantify the history of the feedback in NGC~6338. Qualitatively, we find a set of older lobes connected with the AGN in the southern BCG, and there are also strong indications of a smaller set of lobes from the integrated spectrum and the high spatial resolution radio data. These smaller lobes are expected to be younger due to the slightly flatter spectrum ($\alpha=-0.9$ vs $-1.3$ for the larger lobes), and are orthogonal to the larger lobes in the projection on the sky. The projected direction of the younger lobes agrees with the extent seen in the VLBA image, which could indicate that these are still powered by the current AGN feedback cycle. 
From the spectral shape, the flux, and the three dimensional shape of the outer lobes, where the latter is assumed to be cylindrical, we are able to derive the magnetic field strength assuming energy equipartition (see e.g., \citealp{Govoni2004-dx}).
We use the spectral index and the outer lobe brightness from our spectral modelling (Fig. \ref{fig:spectrum} right), which yields $B=\SI{3.1(5)}{\mu G}$. This magnetic field is not well constrained, mainly due to the uncertainties in the break frequency, but the strength is typical for clusters and groups (\citealp{Dolag2008-bb,Donnert2018-an}). 

The outer lobes are spatially resolved in both GMRT bands, allowing us to construct a spectral index map (Fig. \ref{fig:radio}). However, since in-band spectral measurements are not possible with the available GMRT data, we are unable to measure a potential spectral break from just two frequencies. Instead we can infer a spectral break frequency when we make assumptions on the expansion history of the lobes. 
If the lobe-emission is locally well described by the JP model (\citealp{Jaffe1973-cs}), a widely used spectral aging model assuming a single injection of electrons, (see also \citealp{Harwood2015-do}), and a constant expansion velocity in the plane of the sky from the central AGN is assumed, the spectral break frequency will be proportional to the inverse of the square of the distance $\nu_{\rm break} \propto d^{-2}$, which has also been demonstrated in \cite{Murgia2002-bf,Murgia2003-tm,Parma2007-sd,Giacintucci2008-wf}. 
However, in the case of a continuous injection of electrons by the jets into the radio lobes, the emission can be modeled by a continuous injection (CI) model, which includes an injection spectral index of the youngest electrons that show up in the low-frequency part of the spectrum, and above the break frequency a second, steeper powerlaw. If the injection occurs in phases as it can be assumed for an AGN duty cycle, the CI$_{\rm OFF}$ model is commonly used  (see \citealp{Murgia2011-wj}), which has an additional parameter for the fraction of time that the source is active.

\begin{figure*}
    \centering
    \includegraphics[width=0.49\textwidth]{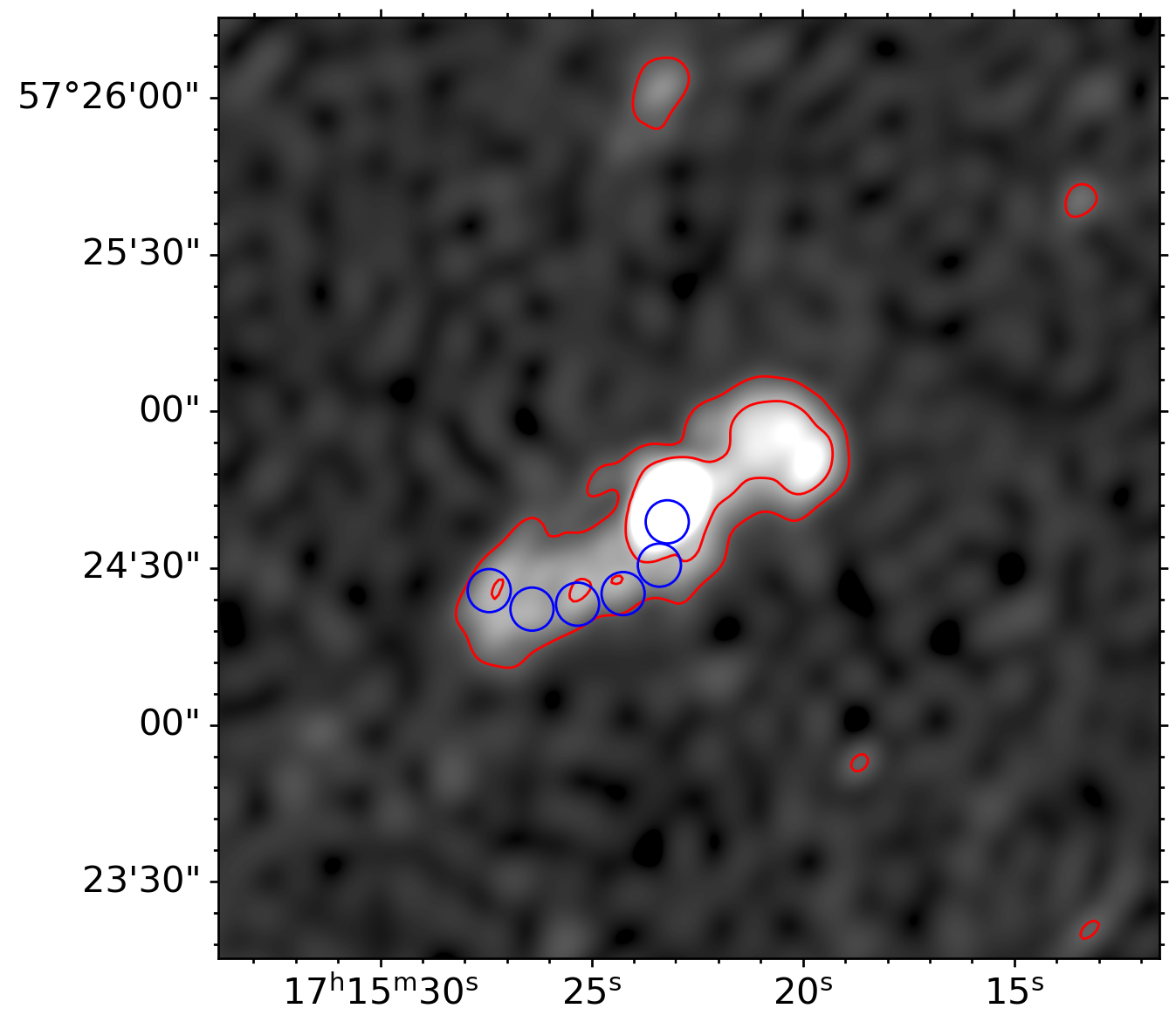}
    \includegraphics[width=0.48\textwidth]{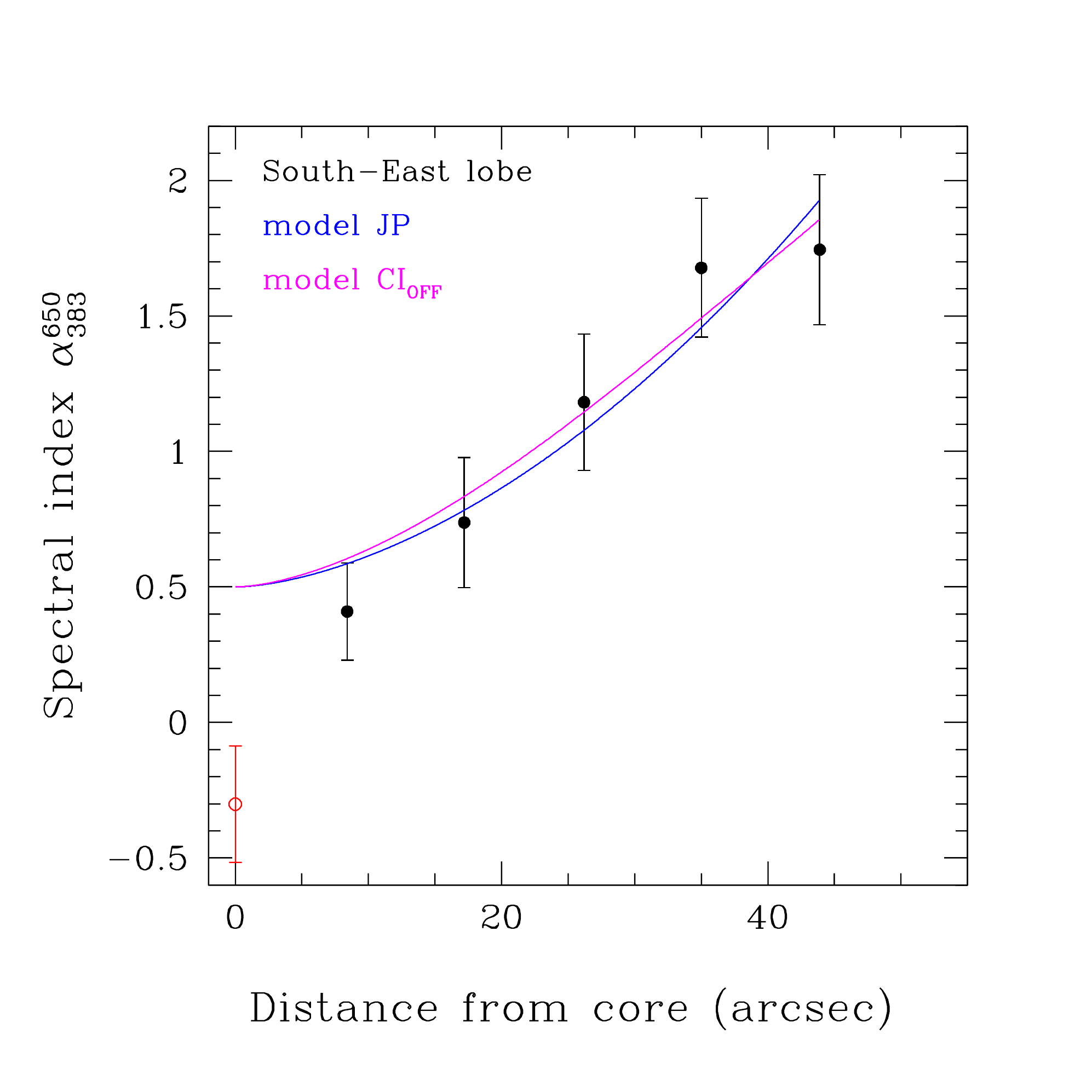}
    \caption{\textit{Left}: GMRT band 3 image with red contours at 5 and 25 $\sigma$ to show the circular regions which were used to extract the spectral index in the two uGMRT bands. \textit{Right:} uGMRT spectral index from 383 to 650\,MHz versus the distance from the core of the southern BCG. This allows to determine the break frequency of the aged spectrum of the lobes.}
    \label{fig:break_trend}
\end{figure*}
We define non-overlapping regions along the lobes centered on the radio brightest part at each distance (see Fig. \ref{fig:break_trend} left) to perform a point-to-point analysis of the spectral index. The regions resemble the circular $\SI{8}{\arcsec}$ beam shape of the reconstructed images that were used for creating the spectral index map. For each region we calculate the flux in the two uGMRT bands, the uncertainties, and the distance to the core along the trajectory, which we show in Fig. \ref{fig:break_trend}. The absolute of the spectral index increases along the south-east lobe, while we do not find a clear trend for the north-west lobe. 
The south-east lobe can be fitted with the JP and the CI$_{\rm OFF}$ models described above using the \verb|Synage++| package. 
Attempting to fit a pure CI model resulted in a very poor fit $\chi_{\rm red}^2 = 4.7$. The JP model  gives a spectral break frequency $\nu_{\rm break}=426^{+164}_{-60}\,\si{MHz}$, while the CI$_{\rm OFF}$ model results in a significantly lower break frequency of $\nu_{\rm break}=177^{+115}_{-117}\,\si{MHz}$, and both fits of both models are good $\chi_{\rm red}^2 \approx 1$. For the CI fits we used an injection spectral index $\alpha_{\rm inj}=-0.5$.

The uncertainties were computed assuming a 10\% flux calibration uncertainty of uGMRT (e.g., \citealp{Kale2022-oi}). 
With the magnetic field strength derived above and the continuous injection model described in \cite{Murgia2002-bf} we are able to derive a radiative age for the lobe of $302^{+144}_{-67}\,\si{Myr}$, while in the JP model we have a slightly lower age of $\SI{197(25)}{Myr}$. 
Assuming a merger velocity in the plane of the sky of $\SI{1050}{km\,s^{-1}}$ (based on the merger velocity of $
\SIrange{1700}{1800}{km\,s^{-1}}$ and a line-of-sight velocity between the BCGs of $\SI{1400}{km\,s^{-1}}$, see \citealp{OSullivan2019-le}), we derive a projected travel distance of 6.5 to 10.2\,arcmin (depending on the model) within the time since the outer lobes were powered. The length of the southern tail is about half that size (\citealp{OSullivan2019-le}). However, we note that the uncertainties on the merger velocity are quite substantial, probably exceeding 20\%. 
The  CI$_{\rm OFF}$ model provides us with further information about the AGN duty cycle: We find that about 44\% of the time the AGN is in the dying phase without injection of new charged particles, although again, the uncertainty on this number is substantial, around 50\%.
If we assume the old and new lobes share the same expansion velocity, we can measure the age of the new lobes by scaling the previously determined age by the relative size of the new lobes. The distance of the outer edge of the new lobe is about 25\% of the trajectory of the old lobes (see Fig. \ref{fig:cavities}), which means that the age of the new lobes is $\SI{50(18)}{Myr}$. 

\subsection{Current state of the AGN}
\label{ch:discussion_agn}
The central radio source in the southern BCG, responsible for the feedback cycles imprinted in the IGM and extended radio emission, has to be understood to obtain a fully consistent picture of the feedback in a group merger. 
The radio spectrum is a powerlaw over almost 3 orders of magnitude, and the spectral index of $\alpha=-0.58$ can be confirmed  through our recent SMA observation at 225\,GHz. 
A flux increase at mm wavelengths was indicated through the 2012 IRAM/GISMO observations, which could not be confirmed with our recent SMA observation. A luminous ADAF component, which could explain such a flux increase could therefore not be confirmed.  The time variability of the $\sim$GHz flux found in the VLA observations (Fig. \ref{fig:lightcurve}), roughly coincides with the time of IRAM/GISMO, which could point to an ongoing feeding/feedback process.

Two absorption mechanisms are typically discussed in literature (e.g., \citealp{Kellermann1969-zj,ODea1998-zj,ODea2021-rn}), free-free absorption (FFA) in an ionized gas surrounding the source, and self-absorption of the synchrotron emitting electrons (synchrotron self-absorption, SSA). Both processes cause an inverted spectrum at low frequencies. In the case of SSA, the theoretically expected spectral index of 2.5 is rarely observed. 
The SSA mechanism requires magnetic fields in the center of the radio source roughly consistent with minimum pressure (\citealp{ODea1998-zj}), which makes SSA often a preferred mechanism, while FFA requires very high thermal electron densities.

It remains difficult to distinguish between FFA and SSA for peaked spectrum sources (e.g., \citealp{Snellen2000-yf,Edwards2004-ih,Callingham2017-uf,Ross2021-ex}).
At lower radio frequencies (around 800\,MHz) we find a turnover of the spectrum, from a spectral index of $-0.58$ above 1\,GHz toward $>+1.5$ at lowest frequencies. This approaches the homogeneous, opaque synchrotron source spectrum of $2.5$. 

The VLBA image in Fig. \ref{fig:vlba} shows only two components, one of which is assumed to be a jet, the other one likely the core. The visible jet is likely inclined toward the observer, making the current jet system unaligned with the plane of the sky. Since we are unable to distinguish which component is the jet and which is the core, we cannot state the orientation. However, the older lobes in the uGMRT image (Fig. \ref{fig:cavities}) appear more similar in brightness, supporting the assumption that the older outburst did not have a strong line-of-sight component. 

\section{Summary}
\label{ch:summary}
NGC~6338 is one of the most violent group-group mergers known to date. We have examined the past and ongoing feedback processes induced by the AGN in the southern BCG. Interestingly the feedback does not seem to stop, but its direction, traced by the radio lobes and jets, has changed. While the northern group is slightly less massive, hosts also a less massive BCG with weak signs of feedback, the southern BCG provides many features to be explored. 
We find 
\begin{itemize}
    \item a set of old and new lobes (radiative age of $\SI{200}{Myr}$ and $\SI{50}{Myr}$) which seem to be coincide with X-ray cavities,
    \item a change in projected alignment of the two feedback cycles, which is almost $90^\circ$,
    \item an SED from 144\,MHz to 235\,GHz that can be modeled with a powerlaw for each sets of lobes and a powerlaw with a soft turn-over representing the core emission,
    \item larger SE-NW lobes that have a slightly steeper spectral index than the smaller SW-NE lobes ($-1.3$ vs. $-1.1$), and the spectral index steepening of the larger lobes points toward a radiative age of about 200\,Myr, 
    \item a small, parsec-scale jet in the VLBA image, with a direction similar to the new radio lobes, which strengthen the assumption of a recent shift in feedback direction, and that the newly inflated cavities might still be powered,
    \item radio emission in the core that is consistent with typical spectral index of $-0.58$, and shows currently no indications of steepening.
\end{itemize}
For a complete picture of the feedback processes, especially on the galaxy group level, we need to understand the current and past AGN feedback cycles in various environments.
Through the wealth of high-quality, multiwavelength data from radio to X-rays, NGC~6338 offers the possibility to look at the feedback process in a violent group-group merger in the pre-core-passage phase.

The results highlight the importance of a good understanding of the radio/mm SED, and reveal another case with a recent change in the feedback directional axis.

A future, high spectral resolution X-ray imaging instrument such as Athena X-IFU (\citealp{Barret2013-hy}) will be able to provide significant insights on the gas cooling near the center of this group. 

\appendix
\section{Spectral index error map}
\begin{figure}[hb]
    \centering
    \includegraphics[width=0.5\textwidth]{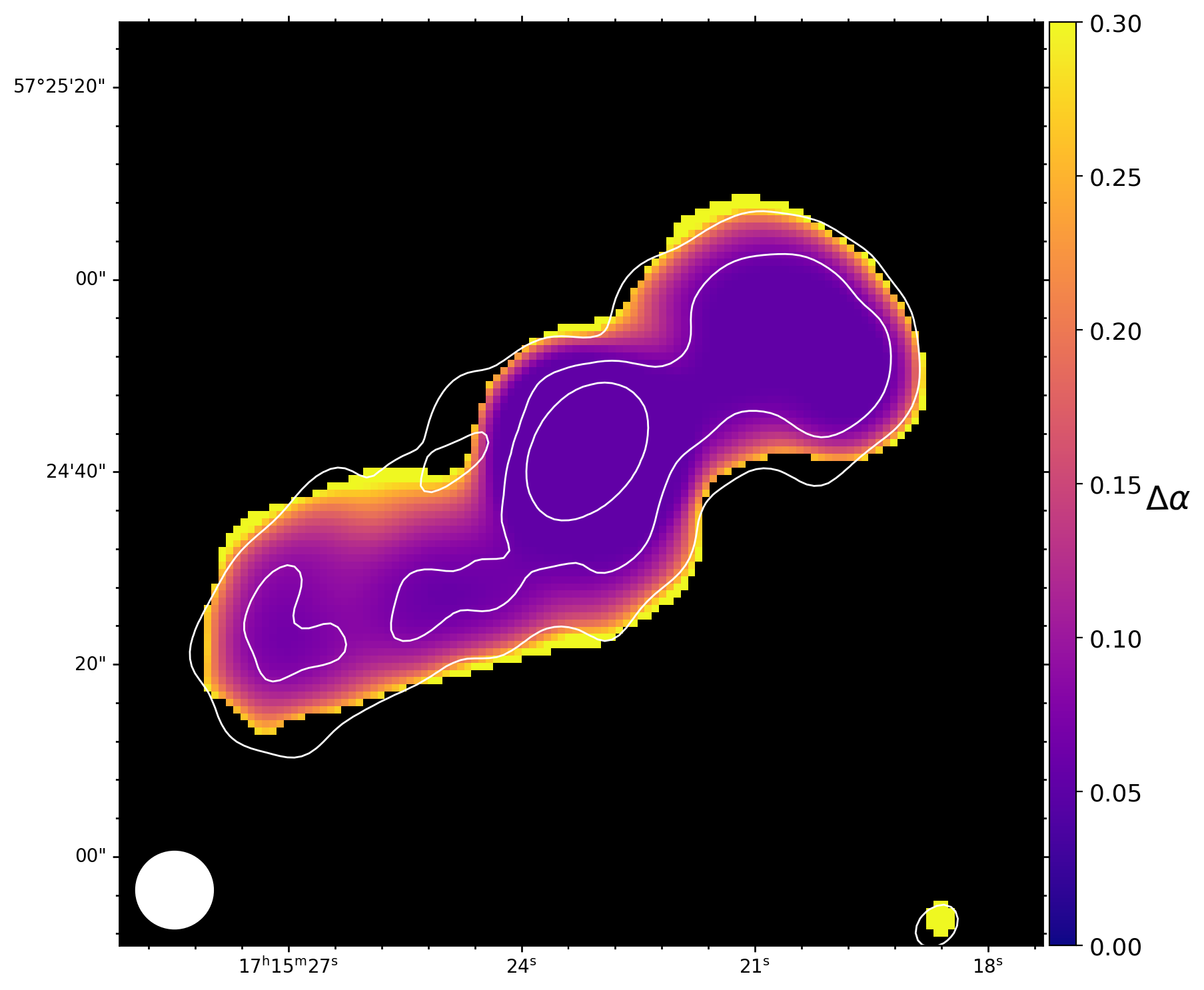}
    \caption{Spectral index error map of uGMRT band 3 and 4 data. The white contours show the emission at 383\,MHz at 5, 25, 125 $\sigma$. The restoring beam is shown in the lower left corner.}
    \label{fig:spixerr}
\end{figure}
\section{LOFAR image}
\begin{figure}[hb]
    \centering
    \includegraphics[width=0.45\textwidth]{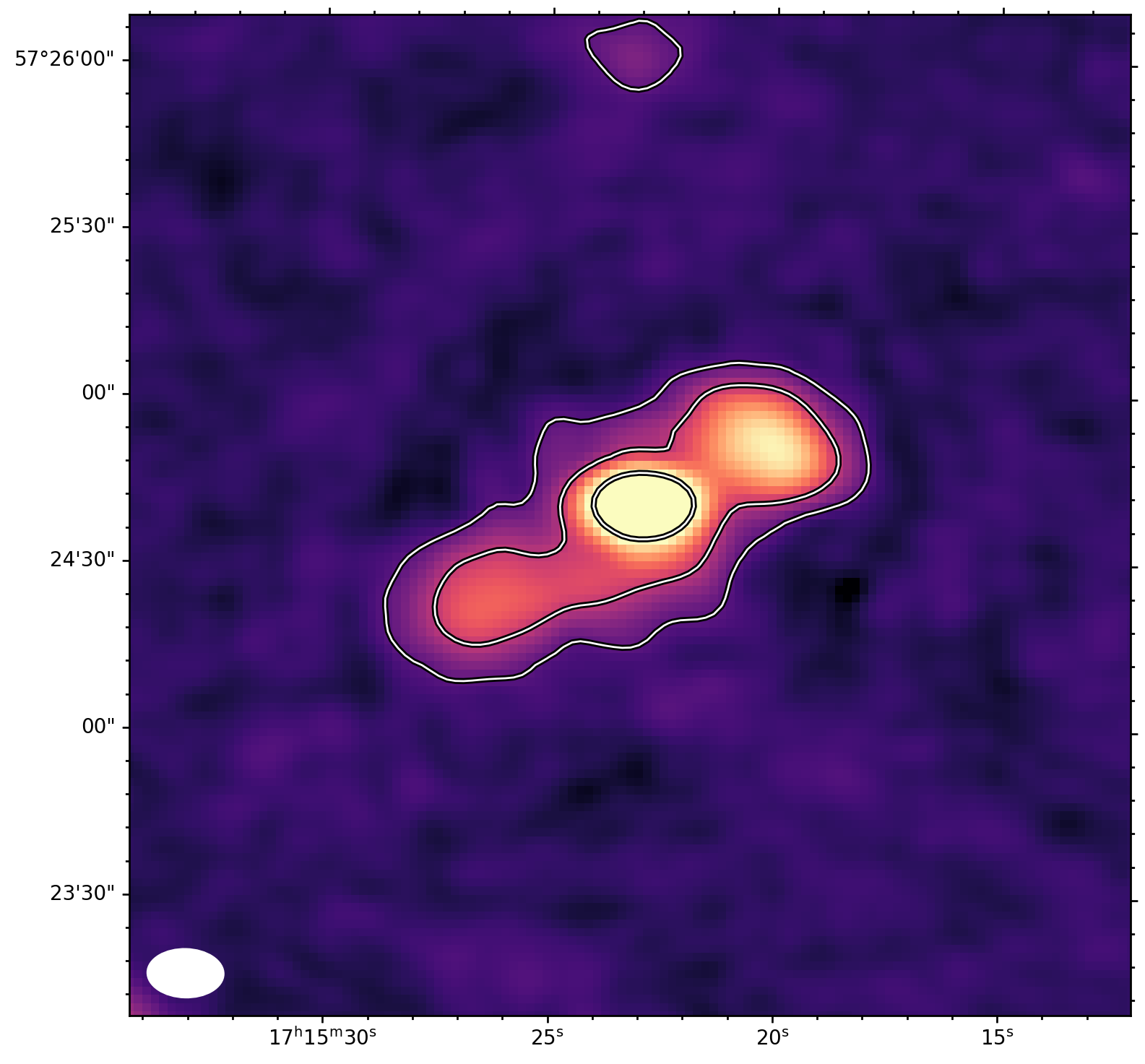}
    \caption{LOFAR image of the southern BCG at $\SI{143}{MHz}$ with contours at 5, 25, 125 $\sigma$, see also \citealp{Birzan2020-oe}. The shortest LOFAR baselines are 90m at $\SI{140}{MHz}$, corresponding to a largest angular scale of $\SI{1.3}{\deg}$. However, the LOFAR image does not show any emission on scales larger than 1.5arcmin.}
    \label{fig:lofar}
\end{figure}

\clearpage
\begin{acknowledgments}
GS acknowledges support through Chandra grants AR9-20013X. Basic research in radio astronomy at the Naval Research Laboratory is supported by 6.1 Base funding. 
EOS acknowledges support through Chandra grant G08-19112X. 
We thank the staff of the GMRT who have made these observations possible. The GMRT is run by the National Centre for Radio Astrophysics of the Tata Institute of Fundamental Research. 
HAP acknowledges support by the National Science and Technology Council of Taiwan under grant 110-2112-M-032-020-MY3.
\end{acknowledgments}

\software{
AIPS (\citealp{Greisen1990-ak,Greisen2003-le}),
astropy (\citealp{The_Astropy_Collaboration2013-lw,The_Astropy_Collaboration2018-gx}), 
CASA (v5.6.0 \citealp{McMullin2007-ed}), 
emcee (\citealp{Foreman-Mackey2013-hw}),
MIR (\citealp{Gurwell2007-wz}),
PyBDSF (\citealp{Mohan2015-hz}),
SPAM (\citealp{Intema2009-cs})}

\bibliographystyle{aasjournal}
\bibliography{Paperpile_remote.bib}

\end{document}

%% file: table_obs_gmrt.tex
\begin{deluxetable*}{cccccc}
	\tablecaption{Summary of the NGC\,6338 uGMRT observations for project 39\_009 (PI: O'Sullivan). \label{tab:gmrt}}
	\tablehead{
	    \colhead{Date} & \colhead{Correlator} & \colhead{Bandwidth} & \colhead{Time}  & \colhead{r.m.s} & \colhead{Beamsize} \\
		\colhead{} & \colhead{(MHz)} & \colhead{(MHz)} & \colhead{(min)}  & \colhead{($\si{\mu Jy/beam}$)} & \colhead{($\arcsec \times \arcsec$)}  \\
		\colhead{(1)}  &  \colhead{(2)} & \colhead{(3)} & \colhead{(4)} & \colhead{(5)} & \colhead{(6)}
	}
	\startdata
    March 5 2021 & GWB Band 4/650 & 200 & 223 & 16 & $4.4\times 3.3$ \\
    March 12 2021 & GWB Band 3/383 & 167 & 199  & 32 & $7.9\times 5.6$ \\
	\enddata
	\tablecomments{Column 1: Observing date. Column 2: Correlator and observing frequency. Column 3: bandwidth. Column 4: Time on target. Column 5: image r.m.s. level ($1\sigma$). Column 6: restoring beam.}
\end{deluxetable*}


%% file: table_obs_vla.tex
\begin{deluxetable*}{cccccccc}
	\tablecaption{Summary of the VLA observations \label{tab:vla}}
	\tablehead{
		\colhead{Project} & \colhead{Configuration} & \colhead{Date} & \colhead{Band} & \colhead{Time} & \colhead{Calibrator} & \colhead{r.m.s} & \colhead{Source flux} \\
		\colhead{}  &  \colhead{} & \colhead{} & \colhead{} & \colhead{(min)} &\colhead{Flux/Phase} & \colhead{($\si{mJy/bm}$)}&  \colhead{($\si{mJy}$)}  \\
		\colhead{(1)}  &  \colhead{(2)} & \colhead{(3)} & \colhead{(4)} & \colhead{(5)} & \colhead{(6)} & \colhead{(7)}& \colhead{(8)}
	}
	\startdata
    AS452 & CnB & Jan 1992 & C & 5.2 & 3C286/1658+476 & 0.4 & 30.9\\
    AF233 & A & Oct 1992 & C & 1.0 & 3C286/1638+573 & 0.59 & 31.4\\
    AC308/NVSS & D & March 1995 & L & 0.5 & 3C286  & 0.45 & 57.5\\
    AE110 & C & June 1997 & C & 9.8 & 1635+381/1740+521 & 0.09 & 27.9\\
    AB879/FIRST & B & July 1998 & L & 3.0 & 3C286  & 0.14 & 49.1\\
    AM701 & D & Dec 2001 & C & 1.0   & 3C286/1740+521 & 0.36 & 25.6\\ 
      "    & "  &     "     & X &  "   &      "          & 0.14 & 20.0\\
      "    & "  &     "     & Ku& "    &       "         & 0.42 & 14.3\\
      "    & "  &    "      & K & "    &       "         & 0.42 & 11.7\\
    12A182 & B & June 2012 & C & 14.4 & 3C286/Multiple & 0.018 & 28.1\\
    15A215 & A & Aug 2015 & C & 40.6 & 3C286/1740+521 & 0.013 & 28.1\\
    VLASS T25t13 & B & Oct 2017 & S & 1.0  & 3C286+3C138  & 0.12 & 44.8\\
      "   &  " & Aug 2020 & "  & "  & "  & 0.14 & 42.8
	\enddata
	\tablecomments{Column 1: project code. Column 2: VLA configuration. Column 3: Observing date. Column 4: Observing band. Column 5: Time on target. Column 6: Flux and phase calibrator. Column 7: image r.m.s. level ($1\sigma$). Column 8: NGC6338 source flux. The corresponding frequencies to the VLA bands are: L - 1.5\,GHz, S - 2.3\,GHz, C - 5\,GHz, X - 10\,GHz, Ku - 15\,GHz, K - 23\,GHz.}
\end{deluxetable*}